\documentclass[sigconf]{acmart}
\AtBeginDocument{%
  }

\copyrightyear{2026}
\acmYear{2026}
\setcopyright{cc}
\setcctype{by-nc-nd}
\acmConference[ICSE-SEIS '26]{2026 IEEE/ACM 48th International Conference on Software Engineering}{April 12--18, 2026}{Rio de Janeiro, Brazil}
\acmBooktitle{2026 IEEE/ACM 48th International Conference on Software Engineering (ICSE-SEIS '26), April 12--18, 2026, Rio de Janeiro, Brazil}
\acmPrice{}
\acmDOI{10.1145/3786581.3786931}
\acmISBN{979-8-4007-2424-4/2026/04}

\usepackage{fbox}
\newcommand{\rqsummary}[2]{
        \vspace{2mm}
        \noindent
        \fbox{%
            \parbox{.97\linewidth}{%
                    \textbf{#1 Summary}
                #2
            }%
        }%
        \vspace{2mm}
}%

\usepackage[most]{tcolorbox}
\usepackage{xcolor}

\usepackage[most]{tcolorbox}
\usepackage{xcolor}
\definecolor{nbpurple}{RGB}{156,89,209}

\newtcolorbox{transnbflagbox}[1][]{
    colback=white,
    boxrule=0.5pt,
    top=6pt,
    bottom=5pt,
    left=2pt,
    right=2pt,
    colframe=black,
    enhanced,
    overlay={
        \fill[cyan!40!white] (interior.north west) rectangle ([yshift=-0.2\ht\strutbox]interior.north east);
        \fill[pink!80!white] ([yshift=-0.2\ht\strutbox]interior.north west) rectangle ([yshift=-0.4\ht\strutbox]interior.north east);
        \fill[white] ([yshift=-0.4\ht\strutbox]interior.north west) rectangle ([yshift=-0.6\ht\strutbox]interior.north east);
        \fill[pink!80!white] ([yshift=-0.6\ht\strutbox]interior.north west) rectangle ([yshift=-0.8\ht\strutbox]interior.north east);
        \fill[cyan!40!white] ([yshift=-0.8\ht\strutbox]interior.north west) rectangle ([yshift=-1.0\ht\strutbox]interior.north east);

        \fill[black] (interior.south west) rectangle ([yshift=0.2\ht\strutbox]interior.south east);
        \fill[nbpurple] ([yshift=0.2\ht\strutbox]interior.south west) rectangle ([yshift=0.4\ht\strutbox]interior.south east);
        \fill[white] ([yshift=0.4\ht\strutbox]interior.south west) rectangle ([yshift=0.6\ht\strutbox]interior.south east);
        \fill[yellow] ([yshift=0.6\ht\strutbox]interior.south west) rectangle ([yshift=0.8\ht\strutbox]interior.south east);
    },
    #1
}

\newcommand{\rqlearning}[2]{
    \vspace{0.4em}
    \begin{transnbflagbox}
        \textbf{#1} #2
    \end{transnbflagbox}
}

\usepackage{cleveref}

\begin{document}

\title[Beyond the Binary: Transgender and Non-binary Software Engineering Students]{Beyond the Binary: Motivations, Challenges, and Strategies of Transgender and Non-binary Software Engineering Students}

\author{Isabella Gra{\ss}l}
\email{isabella.grassl@tu-darmstadt.de}
\orcid{0000-0001-5522-7737}
\affiliation{%
  \institution{Technical University of Darmstadt}
  \city{Darmstadt}
  \country{Germany}}

\begin{abstract}
When software is designed by people from diverse identities and experiences, it is more likely to be inclusive and address a broader range of user needs. 
However, for transgender and non-binary students in software engineering, the path to becoming such creators may be marked by unique challenges. While existing research explores gender minorities in professional software engineering, limited attention has been given to their educational journey, a key phase for ensuring equal opportunities and preventing exclusion in the tech workforce.
This study aims to address this gap by investigating the experiences of transgender and non-binary students in software engineering, with a particular focus on their motivations for entering the field, the obstacles they encounter, and potential strategies for fostering greater inclusivity within their academic environments.
Based on 13 semi-structured interviews with transgender and non-binary students across the globe, we found that gender identity plays an indirect role in their decision to pursue software engineering. Key factors include the appeal of remote work and a personal desire to create more inclusive technologies.
Although the participants did not report direct discrimination within their universities, many described experiencing verbal insults, judgment, intolerance, and hostility, all of which negatively impacted their mental health. These challenges often stem from socio-cultural norms and a lack of representation. Despite these obstacles, the students remain committed to their choice of study but call for greater institutional support, structural changes, and increased representation.
From these findings, we suggest concrete steps to support students, regardless of gender identity. 
\end{abstract}

\ccsdesc[500]{Social and professional topics~Gender}
\ccsdesc[500]{Software and its engineering~Software creation and management}
\ccsdesc[500]{Applied computing~Education}

\keywords{Diversity, Gender, Transgender, Software Engineering Education.}

\maketitle

\section{Introduction}
\emph{“I want to create technology that makes spaces more inclusive.”} (P02)\footnote{\emph{P} followed by a number represents a participant ID from our study. This quote is from a non-binary software engineering student. Non-binary individuals do not exclusively identify with their assigned birth sex and may have a gender identity that falls somewhere along a spectrum or be entirely different from female or male~\cite{americanpsychologicalassociation2018,beischel2023}.}

The desire to create more inclusive spaces is particularly relevant in software engineering, a sociotechnical field~\cite{groeneveld2022a} that extends beyond technical skills to include teamwork and shaping technologies that impact society~\cite{desouzasantosDiversitySoftwareEngineering2023,carver2021,storey2020,staron2024a}. 
When software is designed by individuals and teams from a range of identities and experiences, it is more likely to address diverse needs and perspectives~\cite{canedo2021c, blincoe2019b,catolino2019,groeneveld2022a,albusaysDiversityCrisisSoftware2021,staron2024a}. 
However, for transgender\footnote{Trans individuals have a gender identity that differs from their assigned birth sex~\cite{ashley2024}.} and non-binary students in software engineering, the path to becoming such creators can be marked by challenges~\cite{reggianiLGBTAcademicsPhD2023,mara2021strategies,earle2024a}.

Although gender identity might seem personal, it frequently intersects with one’s professional life~\cite{mara2021strategies,kosciw2015}, particularly in fields like software engineering, which have been dominated by cisgender, heterosexual men~\cite{desouzasantos2024,canedo2023} and shaped by a prevalent \emph{dude} and \emph{bro culture}~\cite{miller2021,kohl2024}. 
This culture can result in feelings of isolation, a lack of belonging, and both overt and subtle discrimination against those who do not conform to traditional gender norms~\cite{stoutLesbianGayBisexual2016a,mara2021strategies,lombardi2002,desouzasantos2023}. Discrimination undermines the collaborative nature of software engineering, where teamwork and diverse perspectives are critical for success~\cite{damian2024,earle2024a,fitzgerald-russell2022}.
These dynamics affect students' personal experiences~\cite{fitzgerald-russell2022} and influence the broader educational and professional environment~\cite{pournaghshbandPromotingDiversityInclusiveComputer2020}, as institutions must ensure they do not disadvantage any group.

While many diversity efforts in software engineering education focus on increasing the representation of women, other gender minorities\footnote{Gender minorities refer to individuals whose gender identity or expression differs from the societal norms associated with the female and male binary~\cite{americanpsychologicalassociation2018,beischel2023}.} are often overlooked~\cite{menierBroadeningGenderComputing2021, casper2022revealing,santosLGBTQIAVisibilityComputer2023}. To address the ongoing \emph{diversity crisis}~\cite{albusaysDiversityCrisisSoftware2021}, efforts must expand beyond the binary and encompass a wider spectrum of gender identities~\cite {begoDiversityInclusionEngineering2021}.
Although some studies have examined the challenges faced by transgender and non-binary software professionals~\cite{desouzasantos2023,desouzasantos2024,desouzasantosDiversitySoftwareEngineering2023,ford2019c,desouzasantosWhatTransgenderSoftware2023}, there remains a significant gap in understanding their experiences as students~\cite{rodriguez-perez2021d,prado2020,santosLGBTQIAVisibilityComputer2023,pradell2024}. 
Most research focuses on the \emph{LGB monolith}~\cite{jennings2020}, i.e., mostly white, cisgender\footnote{This refers to people whose gender identity aligns with their assigned birth sex~\cite{ashley2024}.}, middle-class, and homosexual men and women, often overlooking transgender and non-binary individuals due to their small numbers or treating them as statistical outliers~\cite{jennings2020,desouzasantos2024}. This limits a nuanced understanding of the specific experiences of transgender and non-binary students.

This study explores challenges faced by transgender and non-binary students in software engineering and proposes actionable steps to create a more inclusive academic environment.

We conducted 13\footnote{We acknowledge our small cohort, but recruiting transgender and non-binary students in software engineering is highly challenging~\cite{desouzasantos2024,ellard-gray2015}. Participation requires trust and willingness to share personal experiences, and our aim is not to be fully representative but to give these students a voice. We hope this work inspires further research and encourages others to share their perspectives.} semi-structured interviews with transgender and non-binary students from diverse backgrounds and institutions.
We first examine whether their gender identity shapes their motivation to pursue software engineering.

\textbf{RQ1:} \emph{What motivates transgender and non-binary students to pursue software engineering as a field of study and career?}

Once in the field, these students may encounter obstacles, reflecting the experiences gender minorities face in daily life. We aim to understand the specific barriers within software engineering.

\textbf{RQ2:} \emph{What challenges do transgender and non-binary students encounter in software engineering education?}

Finally, we seek ways to improve inclusivity within the field.

\textbf{RQ3:} \emph{What strategies can be implemented to create a more inclusive environment for transgender and non-binary students?}

Our findings reveal that while gender identity was not always the primary motivation for choosing software engineering, many students were driven by a desire to build more inclusive technologies, shaped by their own experiences of marginalisation. Remote work opportunities and strong career prospects also supported their persistence despite challenges.
Though direct discrimination at university was rare, all participants reported indirect hostility, verbal insults, and judgment linked to cultural norms and a lack of representation, which harmed their mental well-being. 

To foster inclusivity, institutions should increase representation of diverse gender identities, strengthen peer support networks, and address exclusionary cultural norms. Such steps aim to enhance diversity and innovation in software engineering.

\section{Background and Related Work}
Gender identities are complex and often controversial~\cite{ashley2024}. While \emph{sex} and \emph{gender} are frequently used interchangeably, they refer to different aspects of identity. 
Sex denotes biological characteristics~\cite{beischel2023}, whereas gender refers to socially constructed roles, norms, behaviours, and identities~\cite{beischel2023}. 
While many individuals align with their sex assigned at birth, gender minorities do not conform to these or their associated societal expectations, including transgender, non-binary, and gender non-conforming individuals~\cite{beischel2023}.\footnote{In the U.S, over  2.8 million adults and youth identify as transgender~\cite{kidd2021,herman2025}.} 

\subsection{Overview of Gender Diversity}
\emph{Transgender} individuals are those whose gender identity differs from the sex they were assigned at birth~\cite{ashley2024}. This group includes people who transition from male to female (\emph{trans women}), female to male (\emph{trans men}), or who adopt gender identities that do not fit within the traditional binary~\cite{ashley2024}. 

\emph{Non-binary} individuals identify outside the male-female gender binary. This broad group includes gender identities such as genderqueer, genderfluid, and agender. 

\emph{Genderfluid} refers to a gender identity that shifts over time, with a person feeling aligned with different genders at different times.

\emph{Gender non-conforming} individuals may not identify with traditional gender norms, regardless of their gender identity.

\emph{Diversity} within academic environments can be understood along two dimensions~\cite{gardenswartz2003}, with internal diversity encompassing factors such as age, ethnicity, gender, and sexual orientation. External diversity includes factors such as education level, socioeconomic status, and work experience. 
For transgender and non-binary students, their internal diversity, in terms of gender identity and expression\footnote{Gender expression is how an individual outwardly shows their gender identity through clothing, behaviour, and appearance.}, profoundly shapes their experiences, often influencing how they are perceived and treated by others~\cite{pournaghshbandPromotingDiversityInclusiveComputer2020}. 

In general, studies have highlighted that transgender and non-heterosexual students face difficulties, including harassment, and perceived lack of support~\cite{ellis2009diversity,gonzalez2024,lombardi2002, casey2019}. Moreover, especially transgender students often sought safety, and frequently reported feeling isolated or marginalised due to their gender identity~\cite{goldbergWantBeSafe2021}. Mentoring and fostering a sense of belonging are often beneficial in helping these students navigate their academic careers~\cite{sarna2021}. 
These challenges especially create barriers to entry and retention in fields like software engineering, where cisgender norms are dominant, and a lack of understanding can lead to exclusion~\cite{miller2021,canedo2023}.

\subsection{Gender in Software Engineering}
Gender disparities in software engineering persist~\cite{cheryan2024}, with most research focusing on cisgender women. 
However, the experiences of transgender and non-binary individuals have received limited attention, despite their challenges may differ~\cite{vanbreukelen2023a,cheryan2024, santosLGBTQIAVisibilityComputer2023,jennings2020}.
While these studies are valuable, much of the gender-diverse research in software engineering focuses on the broader LGBTQIA+\footnote{LGBTQIA+ is an umbrella term that includes lesbian, gay, bisexual, transgender, queer, intersex, and asexual identities. The \emph{+} represents other non-heteronormative identities.} community, with limited attention paid to the specific gender minorities within this group~\cite{stoutLesbianGayBisexual2016a, santosLGBTQIAVisibilityComputer2023,vorderwulbeke2025}. 

Most studies focused on remote work where professional software developers value the safety offered by remote work, yet it can also lead to increased isolation and invisibility~\cite{desouzasantosWhatTransgenderSoftware2023, desouzasantos2023}. Research has shown that both \emph{older} women and non-binary individuals in software engineering often feel pressure to conform to masculine norms~\cite{vanbreukelen2023a}, and many male developers remain unaware of these inequalities~\cite{canedo2023}.
Efforts have been made to improve LGBTQIA+ representation in the broader software community, but under-representation of non-cisgender identities remains~\cite {zolduoarrati2024}. 

\subsection{Gender in Software Engineering Education}
Most studies target students either only in computing fields, not software engineering, and/or targeting the LGBTQIA+ group as a whole, not explicitly mentioning transgender or non-binary students~\cite{vorderwulbeke2025}.
In generic computing, the lack of visibility is, besides the lack of support and communities, the major issue reported by predominantly non-heterosexual, cisgender engineering students~\cite{whitley2022,boudreau2018,yang2021,sona2023,earle2024a,rosemarypradell2024,bilimoria2009}.
In addition, there are indicators that transgender and non-binary students in computing fields experience higher levels of stress, harassment, and exclusion compared to their cisgender peers \cite{bilimoria2009,nolan2024,sona2023,reggianiLGBTAcademicsPhD2023,whitley2022}. 
The perceived \emph{chilly} \cite{brinkworth2016chilly,atherton2016physics} institutional climate in computer science negatively affects the students' academic performance \cite{garvey2018impact}. 
Microaggressions further exacerbate these negative experiences~\cite{fitzgerald-russell2022}. Transgender and gender non-conforming students in STEM overall are more likely to leave their programs due to hostile or unsupportive environments~\cite{maloyFactorsInfluencingRetention2022}. 



Yet, a gap remains in understanding the experiences of transgender and non-binary students in software engineering~\cite{sulimani-aidan2024,cheryan2024}. As software engineering education prepares students for the professional world, it is essential to tackle gender inclusion at this level to prevent exclusionary practices from persisting in the workforce.

\section{Methodology}
This exploratory study adopts a qualitative approach, using interviews to capture the experiences of transgender and non-binary students in software engineering. The aim is not statistical generalisation. We aim to give these students a voice~\cite{jourian2017} and provide insights that deepen understanding of gender diversity in software engineering education. 

\subsection{Recruitment}
Our target group was students studying software engineering or majoring in software engineering who self-identify as transgender or non-binary. 
Recruiting participants from such a hidden population~\cite{desouzasantos2024} posed challenges, as individuals may not always feel safe or willing to disclose their gender identity due to the sensitive and potentially distressing nature of the subject, as well as past negative experiences.
To overcome this challenge, we employed a multi-pronged sampling strategy~\cite{desouzasantos2023,ellard-gray2015}.

\textbf{Convenience Sampling:} Initially, we reached out to individuals within our personal and social networks who identified with the target group. These initial contacts facilitated participation in a familiar and trusted setting. Of the nine people approached, four participants (P01, P11, P12, P13) agreed to share their experiences.

\textbf{Purposive Sampling:} Recognising that personal and social media recruitment would not yield additional participants, we used the \emph{Prolific} platform\footnote{\url{https://www.prolific.com/}}, which has been effectively used in previous software engineering research~\cite{russo2022b,baskararajah2021}. Prolific enabled us to reach a broader, more diverse pool of students from various countries. Screening questions ensured that only students in software engineering who identified as transgender or non-binary were eligible. Of 251,721 active users, 57 met our criteria, yet only nine responded within five months, even after increasing the compensation rate to an average of \pounds 34 per hour, underscoring the significant recruitment challenges. These nine participants (P02–P10) were included in the study.

\subsection{Participants}
\Cref{tab:demos} provides the demographic information for all our participants. Participants' identities spanned the gender spectrum, from transgender women and men to non-binary and gender-fluid individuals, with varying degrees of gender expression, such as masculine, feminine, or androgynous. All of our participants are \emph{out}, i.e., they express and live in accordance with their gender identity.

Participants came from a range of geographic locations, with seven from Europe, three from North America, one from South America, one from Africa, and one from Asia. This diversity allowed us to capture a wide range of experiences in different cultural and institutional contexts.
The majority of participants ($N=9$) are undergraduate students (Bachelor level) studying software engineering and computer science, with a major in software engineering/programming languages. Participants P12 and P13 are in graduate studies (Master's level), while P09 and P11 are pursuing PhDs in computer science, with a focus on software engineering.\footnote{In most countries, PhDs are awarded in computer or natural sciences, with software engineering pursued as a research focus.}

\begin{table}[t]
		\centering
		\caption{Demographics of all participants.}
		\label{tab:demos}
		\small
		\begin{tabular}{lp{2.3cm}p{1.2cm}p{2cm}p{1cm}}
			\toprule
			ID & Gender Identity & Pronouns &Gender Expression & Country \\ \midrule

			P01  &  transgender woman & she/they  & feminine/ androgynous & Turkey \\  

			P02 &  non-binary & they/them* & androgynous & Greece \\  

			P03  & non-binary, transgender & they/them & 50/50 masculine-feminine & UK \\  

			P04 & no label & he/they  & shift of masculine/feminine & France \\  

			P05 &   non-binary & they/them&  50/50 masculine/feminine & Mexico \\  

			P06 &  genderfluid** & they/she/he  & androgynous & Italy  \\  

			P07 & non-binary & he/him & feminine & USA \\  

			P08 & non-binary & none preferred  & depends on & South Africa \\   

			P09  & transgender woman & she/her  & feminine & Japan\\  

			P10  & transgender man & he/him  & masculine & USA \\  

            P11 &  transgender woman & she/they & feminine & Europe***\\  

            P12 & transgender woman & she/her  & feminine & Canada \\  

			P13  & transgender woman & she/they  & feminine & UK  \\ 
            \bottomrule
		\end{tabular}
		\footnotesize{*Participant prefers these pronouns in English. They stated to use male pronouns in their native language.\\  **Participant stated to use non-binary label when they feel it would be understood more easily, but prefers genderfluid. \\  ***Participant preferred not to publicly share their country.}
\end{table}

\subsection{Interview Process}
All interviews were conducted remotely via the video conferencing platform \emph{Zoom} from July to December 2024. Prior to the interview, participants were asked to provide demographic information, including their country of study and the number of terms they had completed in their program.

The interviews followed a semi-structured format, allowing for flexibility during the conversation. This structure ensured consistency across interviews by using questions that aligned with our three research questions, while also enabling follow-up questions tailored to each participant's experiences and responses, especially regarding their gender identity.\footnote{\url{https://figshare.com/s/296ae9456d9d27df3a51}}
One researcher conducted nine interviews, while another researcher conducted the remaining four. The interviews lasted between 60 and 90 minutes.

\subsection{Data Analysis}
Interviews were audio-recorded with participants' consent and transcribed orthographically using the automatic transcription feature in Microsoft Word 365. Transcriptions included all spoken words and sounds, such as hesitations, speech cut-offs, giggles, long pauses, and strong emphasis, and were then manually reviewed and corrected by two researchers for accuracy.

We employed thematic analysis~\cite{braun2012} to identify, code, and categorise patterns (\emph{themes})~\cite{neuendorfContentAnalysisThematic2018}. A mixed approach was adopted: deductive coding based on our research questions, followed by inductive theme development grounded in the interview~\cite{braun2012}.

\emph{Phase 1} involved familiarisation with the data, including writing analytical memos after each interview and individually reflecting on transcripts to question assumptions~\cite{braun2012}.

\emph{Phase 2} involved systematic coding~\cite{braun2012}. We followed a two-phase coding procedure: training and reliability assessment~\cite{oconnor2020,syed2015,tracy2010}.
Two researchers first discussed the research questions to align understanding, then independently coded two randomly selected interviews. Cohen’s Kappa~\cite{mchugh2012} indicated substantial agreement ($\kappa$=0.72) for the first pair and almost perfect agreement ($\kappa = 0.84$) for a second pair~\cite{landis1977}. After refining the coding approach, the remaining interviews were coded independently by one researcher.

\emph{Phase 3} focused on generating themes from the codes, uncovering shared meanings and constructing logically coherent themes, reflecting the significance of the participants’ experiences~\cite{braun2012}.

We reached saturation for the overarching themes within eleven interviews. However, saturation was not achieved for individual sub-themes, particularly those related to cultural differences.

\subsection{Ethics}
This study adhered to the ethical guidelines set by the Technical University of Darmstadt and standard protocols for conducting research with human participants, particularly vulnerable populations~\cite{vinson2008,ellard-gray2015}. Participants were informed about the purpose of the study and consented to the recording and transcription of their interviews.
To protect participants' privacy, no personal identifying information, such as names or contact details, was recorded. Participants could withdraw from the study at any time.

\subsection{Threats to Validity}
We address the following potential threats to validity~\cite{wohlin2012,ralph2021}.

\emph{Internal Validity} 
A primary limitation is the small sample size of 13 participants, which restricts generalisability and may introduce sampling bias. Recruiting from this hidden population was highly challenging, and our aim is not statistical representation but to give voice to transgender and non-binary students in software engineering. Despite its limits, the study offers rich insights into under-represented experiences. 
Another threat is interviewer bias, which could influence data interpretation. To mitigate this, multiple researchers were involved, and inter-coder reliability was assessed to ensure objective analysis. The research team identifies as cisgender individuals who belong to queer and other under-represented groups; we recognise this positionality as a strength in raising awareness of diversity issues and a limitation in potentially shaping the interpretation of the data.

\emph{External Validity}
Due to the small and specific sample, findings may not be fully generalizable to all gender minority students in software engineering. The experiences of students from particular regions or institutions may differ from those of students in other contexts, influenced by varying cultural norms, institutional structures, and policies. Additionally, while we made efforts to gather data from a variety of participants, some regions or populations may still be under-represented. Therefore, we acknowledge that further research involving larger and more diverse samples across different regions and educational systems is necessary to enhance the external validity of these findings.

\emph{Construct Validity} 
A potential threat is the misinterpretation of interview questions or participants' responses. Given the sensitive nature of the topic, participants may have difficulty fully expressing their experiences. The evolving nature of gender identity may also influence participants' understanding and expression of their experiences over time.
We mitigated these risks by using a semi-structured interview format, which allowed for flexibility and follow-up questions tailored to the individual’s experiences. This ensured that participants had the space to elaborate fully and clarified the context of questions when necessary.
We share the recruitment messages, interview scripts, and final codebook.\footnote{\url{https://figshare.com/s/296ae9456d9d27df3a51}}

\section{Results}
We conducted interviews with 13 transgender and non-binary students (\Cref{tab:demos}) and applied thematic analysis to identify themes that address our research questions.

\subsection{RQ1: Motivation for Studies and Career}
We examined the themes that emerged from codes related to the students' motivations for pursuing studies and a career in software engineering. Each subsection represents a theme. 
While gender identity was not the primary motivation for pursuing software engineering, it influenced \emph{all} participants to varying degrees in their study choice, as one transgender woman exemplified \emph{“I wouldn't say it's the main reason, but it is one of the most important reasons. I can't say I wouldn't be interested into programming if I weren't a transgender woman, but I would be a little bit more confused. It made my decision easier.”} (P01)

\subsubsection{Remote Opportunities: Safety and Self-Expression}
Overall, remote opportunities emerged as a critical factor for both emotional and physical safety, as well as for self-expression for almost all participants ($n=10$; P01, P02, P03, P05, P06, P07, P08, P09, P10, P13). By reducing exposure to potentially hostile environments, remote work and study offer participants a \emph{safer space}, allowing them to avoid situations where their gender identity might lead to discomfort or discrimination.

\paragraph{Protection from Hostile Environments}
For many participants, the ability to study or work remotely on software projects was key to their decision to pursue software engineering, as it provides a space free from the risks associated with in-person interactions, avoiding the discomfort of physically or emotionally hostile situations.
One non-binary student from Mexico highlighted the importance of remote study in mitigating these risks: \emph{“If I had to go to an actual university, to the building, to the campus, I’m not sure how I would behave. It sounds scary, so, it’s easier.”} (P05) 

One non-binary student from Greece echoed this sentiment: \emph{“It’s easier to be nonconforming in gender identities in this space, with people working remotely on their software and interacting remotely; it gives you a safer environment to exist and work.”} (P02)

One transgender woman also expressed the value of remote work for her emotional safety: \emph{“Even though it’s not only because of my gender, my gender expression and my gender identity make it sometimes hard and overwhelming for me to just go outside, and I prefer to stay home. I feel like the kind of job I am looking for may help me stay home and not have to go to an office where I would have to deal with these emotionally draining situations.”} (P13)

\paragraph{Empowerment and Self-Expression}
In addition to offering safety, remote study and work enable participants to express themselves more authentically, free from the pressures of traditional gender expectations in classroom or office settings. 
The freedom to work from a location of their choosing gives students the autonomy to engage with their studies without the constraints of physical spaces.

One genderfluid student from Italy explained how remote study allows them to avoid uncomfortable gendered assumptions: \emph{“When someone sees me in person, they might assume I am a woman, purely because of that. Many people in my course are men, or at least appear to be. Well, I found a couple of other LGBT people I felt comfortable with. The ones who weren’t kind of dismissed most LGBT topics, which made the in-person space feel uncomfortable for me.”} (P06)

Similarly, one transgender student explained how online courses reduce the need to conform to expectations: \emph{“I do feel like it's easier for me to kind of be who I am and be open about it. Because I am not forced to be in a physical classroom.”} (P05)
This is echoed by one transgender woman from the UK: \emph{“Having the freedom to work when and where I want makes me feel like I have more control over my life, eventually feeling free.”} (P13)

\rqlearning{}{Offering remote or hybrid options can improve retention and attract diverse talents in software engineering.}

\subsubsection{Financial Security and Global Mobility}
Beyond safety and self-expression, financial security emerged as  a significant factor, with many participants ($n=7$; P01, P02, P04, P05, P09, P10, P13) noting that the high earning potential and global demand for software engineers were key motivators, as they eventually also allow them to relocate to more LGBTQIA+ friendly countries.

\paragraph{Financial Stability and Job Security}
Participants highlighted that the future-proof nature of software engineering, combined with its lucrative salaries, provided a sense of financial security that might not be offered in other fields. This is particularly important for individuals who may face discrimination in alternative professions. 

One student emphasised the promising future of software engineering: \emph{“It's like a whole universe. This is the job of the future. I mean now. [...] I think software jobs will never die. So for me, that's how it feels. So definitely, I feel more content about it.”} (P04)

One student also added the current situation in their country (UK): \emph{“Have you seen the current job market? Being a software developer is about the only [...] way to stay immune to this mess.”} (P13)

\paragraph{Migration to More Welcoming and Safer Countries}
Given that certain regions are more LGBTQIA+ friendly than others, and where belonging to this group is not criminalised, having a skill set like software engineering can make it easer to migrate to countries with stronger protections and more inclusive policies.

Even though some participants did not fully enjoy their studies (e.g., P02, P05, P10, P13), they recognised this long-term value. One student, for instance, mentioned that despite a shift in their academic interests, they remained optimistic about future job satisfaction:  \emph{“I don’t particularly enjoy the studies, but I can see myself working in the field and building a career.”} (P05)
Another participant added: \emph{“I want to be able to move to a country where I can live more openly and have the job opportunities to support myself.”} (P13)

\rqlearning{}{The demand for software engineers offers independence and relocation, highlighting the need to actively recruit these groups.}

\subsubsection{Opportunities for Personal and Societal Impact}
Several participants ($n=7$; P02, P03, P06,P08, P10, P12, P13) expressed a strong desire to use their skills to create a positive societal impact by making the software industry more inclusive. Their personal experiences of exclusion or under-representation drive them to create accessible technologies for diverse gender identities.

One student from Italy shared their motivation to help others: \emph{“Just like having the same hardships as me, like, I just want to make things that will help both people in software, and like outside of it. [...] Having easy access to communities and information, which honestly helped me a lot. I just really love the things that I'm able to do with the things I'm learning. I like that I'm able to make things more accessible for people that might have issues.”} (P06)

One student from the UK reflected on how their gender identity influenced their drive to develop inclusive technologies: \emph{“As a non-binary individual my gender identity has influenced my desire to create inclusive technology that considers more inclusive technology spaces... Yes, that motivates me a lot.”} (P03)
Similar aspirations were expressed by a non-binary student from Greece: \emph{“My goal is to develop applications and platforms that are accessible to all users regardless of their gender identity and their ability or their background. So this decision is closely tied to my gender identity, as I experience firsthand the barrier.”}
The chance to create meaningful change was seen as more than compensating for the difficulties encountered along the way: \emph{“Despite the challenges, I am content with my choice. It has given me the skills and the knowledge to make a positive impact in the technology industry.”} (P03)

\rqlearning{}{Tech companies can drive innovation by valuing the perspectives of transgender and non-binary individuals, whose focus on inclusivity can enhance solutions for a broader user base.}

\subsubsection{Sense of Belonging Through Representation}
Several participants ($n=6$, P05, P06, P08, P10, P11, P13) shared that seeing transgender and non-binary professionals in the field boosted their confidence and motivation to pursue their studies.

\paragraph{Visibility of Transgender and Non-Binary Role Models}
Participants highlighted that the presence of successful transgender and non-binary professionals in the tech industry made it easier for them to envision themselves thriving in the field.

One transgender student from the U.S. expressed how seeing other transgender individuals in programming and networking roles made a difference: \emph{“I think [my gender identity influenced my choice] because there are so many trans people who work in programming, who work in networking.”} (P10)

Similarly, a transgender PhD student in software engineering reflected on how their gender identity affected their comfort with their studies, especially when they recognised how welcoming parts of the tech community could be: \emph{“I realised I was trans after I had already signed up for the degree... but [my gender] did influence how comfortable I am with my studies because of how much of the Computer Science community online is a bit trans.”} (P11)

\paragraph{Online Communities as Spaces for Belonging}
Beyond role models, online tech communities provide spaces for transgender and non-binary students to feel connected and supported. These virtual, often international, communities allow students to find like-minded peers without fear of judgment.
One student described how online communities provided support: \emph{“Most of my trans friends who are also into programming have helped me feel more comfortable with the subject. It did make me feel more comfortable searching for resources and seeing other people online discussing the subject. It made me feel less afraid to reach out to online communities, though not as much within personal communities, because there’s a higher risk there. [...] We’re all in different countries, but the programming community still feels more welcoming just knowing they’re there.”} (P06)

\rqlearning{}{Representation matters: visibility, mentorship, and inclusive communities inspire and retain gender minorities.}

The motivators identified in this section are interconnected, as one student summarised this interrelationship well: \emph{“I feel like it's harder when you're a more visible trans person in a regular work department. And so that means I am motivated more to find something that will get me out of poverty because we are pretty poor and gives me a more flexibility as someone who's different, because I feel like there's more acceptance too in tech for different [people].”} (P10)

\rqsummary{RQ1}{Transgender and non-binary students are motivated to pursue software engineering for personal safety through remote work, career opportunities, and the desire to make a positive impact, with their gender identity playing an indirect role.}

\subsection{RQ2: Challenges}
We analysed themes from students’ responses about challenges in studying and working in software engineering. Each subsection presents a distinct theme, ranging from subtle discrimination to safety concerns and the need for support in online communities. 

\subsubsection{Microaggressions and Discrimination}
While none of our participants reported experiencing overt violence or severe incidents of discrimination\footnote{It is important to note that those who may have experienced severe discrimination or violence might have already dropped out of the field or may be unwilling to participate in interviews, leading to the underrepresentation of such experiences in our data.}, almost all of them faced subtle, yet impactful, forms of bias. 
These included judgement, verbal insults, microaggressions\footnote{Microaggressions are subtle, often unintentional comments or actions that convey prejudice or discrimination toward a marginalised group}, indirect discrimination, social exclusion and marginalisation, as well as misgendering.
Although these actions were often indirect, they had a profound effect on the mental well-being and emotional safety of the students.

One transgender student from Turkey encapsulates this experience, explaining: \emph{“I didn't directly face any discrimination, I had more like, 'someone said this'… or like, I don't know, people are discriminating towards the general, showing hatred and disgust to LGBTQIA+ community, not directly at me.”} (P01)

\paragraph{Negative Impact on Mental Health}
The indirect nature of these behaviours does not make them any less harmful. 
Almost all of our participants expressed how these actions affected their mental health and sense of safety, making it difficult for them to fully express themselves. 
One student emphasises the emotional toll, stating: \emph{“A lot of people, like, you can tell that they don't care, but you get the odd look sometimes. [...] I think the closest problem I have in the university, with my identity is being able to be how I want without feeling alone and being scared.”} (P02)

Similarly, one student encountered challenges in the form of microaggressions and misgendering from both peers and faculty members. As they explained: \emph{“I encountered instances of misgendering and microaggressions from both students and also the faculty. Which added an emotional burden to me, I think. Yeah. It's another emotional burden to me.”} (P03)
This indicates that even small actions, like incorrect pronoun use, can create an additional emotional strain for gender minority students.
Another student echoed these sentiments, highlighting the emotional pain: 
\emph{“[...] calling names like you see a genderqueer or someone you don't like the appearance and then throwing names or calling names or this stuff it's kind of... Hurtful.”} (P04)
Such experiences added another layer of difficulty to daily life at university: \emph{“And this bitching about me and giving me these looks made my everyday campus life even more challenging. I’m profoundly overwhelmed by it all the time.”} (P13)

\paragraph{Passing versus Expressing Yourself}
Some of our participants reflected on how they avoid drawing attention to themselves by \emph{passing}\footnote{Passing refers to transgender individuals being perceived as the gender they identify with and/or being perceived as cisgender.}: \emph{“I have been fortunately, like I pass and not a lot of people look at me twice. So I haven't had a lot of trouble."} (P12) 
However, this often comes at a personal cost, as they and others like P03 and P08 noted the discomfort of not being able to \emph{“fully express [them] freely”} (P13). 
One non-binary student from South Africa shared a similar experience of being judged based on appearance, reflecting on how the judgement made them reconsider their self-expression: \emph{“When people start judging and looking and pointing out fingers on how I dress or whatsoever, I sometimes think about just not dressing that way, but then I struggle with hiding myself.”} (P08)

One student adds another layer to this, sharing how mismatched gender cues, such as voice and appearance, can make people and also themselves uncomfortable: \emph{“People judging me based on appearance because my voice says one thing, but my face says something else. A lot of people, they feel uncomfortable.”} (P10)
Social interactions were marked by gendered responses from peers as one  genderfluid student from Italy described: \emph{“A woman or like fem presenting people generally were like supportive. They were like, oh, that's that's cool. Like sometimes we talked about our, I talked about my girlfriend, and they talked about their boyfriends. Guys, were  more... Dismissive. I thought it might be like a cultural issue and like how people were raised, I imagined that. I honestly can't figure out how to feel about that. Uh. I told the people I had to do group projects with.”} (P06)

\rqlearning{}{Social marginalisation, through microaggressions, insults, or judgement, harms academic performance and well-being. }

\subsubsection{Culture-dependent Stigma and Barriers} 
Beyond interpersonal discrimination, for students from regions where gender diversity is not widely accepted, safety concerns were a significant factor influencing their ability to express their gender identity. 

Participants from Mexico, South Africa, and the USA expressed heightened concerns about safety due to growing resistance towards non-heteronormative individuals in certain regions.

Particularly, two participants from Mexico and South Africa described concerns that align with findings from recent studies on challenges faced in the Global South~\cite{mutongoza2024}.

Even in more liberal societies, there are still instances where safety and acceptance are not guaranteed.
One student from the US reflected on the broader societal context, expressing concerns over growing hostility: \emph{“ You know there’s backlash right now in the USA against LGBT people. [...] Safety concerns for my family make me consider moving to another neighborhood, and while my gender identity is not the only reason, it is definitely a factor.”} (P10)

In Japan, one transgender woman felt constrained by societal norms and hesitating to take steps like officially changing her name at the university due to the lack of openness regarding LGBTQIA+ issues. 
\emph{“I haven't taken action on changing my registered name at my institution because in Japan, people are not open about LGBTQIA+ matters. [...] like you wouldn't see queer people, like you wouldn't think about it. It's very hidden. I think that's also a very Japanese, it's very hidden like. You don't talk about it.”} (P09). 

She noted the internal conflict caused by society’s tendency to default to assumptions about cross-dressing: \emph{“Here in Japan is that the default is kind of being a crossdresser. So I never know if I pass or if people think I'm a crossdresser and it's kind of, I mean I know I shouldn't care, but then again we go back to the internalized transphobia and it's like a never ending cycle basically.”} (P09)

Similarly, in Turkey, our participant reflected how societal challenges drive transgender individuals to seek opportunities abroad: \emph{“It’s [...] harder to survive if you are a trans woman, in Turkey at least, and being able to develop software, or at least code, usually opens more doors to move abroad.”} (P01)

Further, the lack of representation further compounds the issue: \emph{“I think the representation would solve most of the issues. From what I see from LinkedIn and… other social media, yeah, I think at least, I don’t know I’m not working in the US, and I don’t know the internal stuff but at least they, they seem like they are trying, in some kind of. But in Turkey we don’t have this, we don’t have any kind of, representation. I just saw some Turkish company doing that, and I was really happy. I think it will be enough, because, I think the main problem is… cis-gender people see us as weirdos in some way. But if we had some representation, like, I am doing the same job with you and I’m doing as… good as you. I think they would be more like “oh, okay.” But yes I think this is the only point I have.”} (P01)

One transgender man from the US recounted a challenging experience during an internship: \emph{“He [the boss of the company] only acknowledged two genders, male or female, and nothing outside of that, which I found nonsense. The company was working on a project in Dubai, which was actually very bad. And in the end of the day, it hurt the relationship between me and him, after even I delivered awesome deliverables to him, but still. Most of the time, we were working remotely, and I didn’t really know the people I was working with. It was just about doing what they needed and getting paid.”} (P07) 
This incident created a tense and unwelcoming working environment, which led to strained professional relationships, despite the quality of work produced.

In contrast, students in more accepting countries like France (P04), Canada (P12), and Europe (P11) expressed fewer concerns about openly expressing their gender identity. 
 
\rqlearning{}{Cultural and regional barriers create uneven experiences, sometimes including safety risks for students.}

\subsubsection{Coping Strategies: Online Communities and Peer Support}
In response to these challenges, many participants turned to online communities or supportive peer groups as a means of coping and finding affirmation. Unlike physical academic environments, these online communities allow students to connect with others facing similar struggles, creating a supportive network that mitigates the challenges of studying in more conservative or less accepting physical environments.

One genderfluid student elaborates on how these spaces helped them embrace their gender identity: \emph{“It wasn't necessarily the university or the degree itself that helped me embrace [my gender identity]; it was specifically my close friend group being all comfortable with these things. [...] Most of my online friends have helped me feel more comfortable with myself and feel more comfortable in spaces where I wouldn't be respected as much for my gender.”} (P06)

Online communities and remote learning environments play a significant role in supporting transgender and non-binary students in software engineering, offering them safer spaces to express their identity and connect with others.	
	
\rqlearning{}{Supportive online communities and peer networks help transgender and non-binary students navigate challenges.}

\rqsummary{RQ2}{Transgender and non-binary software engineering students face challenges from subtle discrimination, such as microaggressions, to cultural barriers limiting gender expression. Societal norms and cultural contexts strongly shape how safe and accepted they feel in academic environments.}

\subsection{RQ3: Strategies for Inclusion}
We analysed themes from students’ responses on strategies for inclusion, with each subsection presenting a distinct theme. Overall, participants suggested increasing representation and visibility, as well as structural changes such as gender-neutral restrooms, inclusive scholarships, and training for faculty and students

\subsubsection{Representation and Visibility}
Representation plays a crucial role in breaking down prejudices and creating a more inclusive space. Participants emphasised the need for greater visibility of gender minority groups to encourage diversity and make individuals from these groups feel welcome.

\paragraph{Inclusive Courses}
One non-binary student from Greece reflected on the impact of inclusive language in course content: \emph{“When a gender example comes up, [my professors] make use of both gendered and neutral pronouns, and not a lot of professors do that. [...] They sometimes give queer examples for some of their exercises. It’s always made me feel more welcomed. [...] Maybe like that was officially implemented through the institution, like call professors would sometimes use more inclusive language to make you feel like you are welcome, that it is a safe space.”} (P02)

One transgender woman in Canada echoed this, stating that gender-neutral examples and inclusive content created a sense of belonging: \emph{“Representation is a huge factor to break down prejudices and reflect the reality of people who belong to gender minorities rather than alienating them.”} (P12) 

However, it is important that efforts towards visibility are made cautiously to avoid putting individuals at risk. Publicizing someone's gender identity can expose them to safety risks, so inclusion efforts must prioritize personal safety and autonomy.

\paragraph{Understanding and Showing Support}
A recurring theme among participants was the desire for greater awareness and consistent support from those outside the gender minority community. 

Being visible and raising awareness is the first step: \emph{“People just aren't aware that we exist, and then you're treated kind of like oh, wow.”} (P10) 
One student shared their positive experience with peer support in online spaces: \emph{“I found support through online communities and my friend group, who are also transgender and studying computer science alongside me.”} (P06)

In particular, participants (e.g, P01, P10, and P13) pointed out that the field of software engineering, compared to areas such as the arts or social sciences, were perceived as less welcoming for individuals from gender minorities~\cite{whitley2022}:
\emph{We had a pride parade, just last month and a few of the professors joined us. But most of them weren’t in Computer Science Department, or at least if I recall correctly, none of them was in the Engineering Department. I think this is kind of the stereotypical, like, social sciences professors joined us or, show some support but computer science teachers don’t do that. But, uh, I think, if they did, it would be more helpful because I think, like, I would love to see some support, and, *shrugs* I would love to see some representation, before that.”} (P01)

In addition to the division of support within academic departments, the same student reflected on the tendency for support to be contingent on visible crises rather than an ongoing commitment to inclusion: 
\emph{“Like when there is a problem against, hate crime is going on against LGBT people I see some people showing their support and that’s really nice but a lot of the time they don’t know that, what we are facing and, though they are not aware of us, we are also facing some discrimination. In those times they are usually like, I don’t know, it feels like they don’t care any about us besides when there is a big problem going on. I think it shouldn’t be like that, like you shouldn’t just care about someone when they are really really in that problem. Like, you should, I think, for example in our pride parade this year I don’t remember seeing anyone, that I know, like I knew the most people there, and I don’t remember someone who’s identifying as cisgender and heterosexual. But last year there were. There were problems and they showed their support. This year they didn’t see any problem, so they didn’t come. I think this is kind of wrong, and I would love to see them join us in these times too.”} (P01)
This observation underscores a deeper issue of conditional support, where the commitment to inclusivity is more reactive than proactive.

\paragraph{Outreach to Younger Audiences}
Participants highlighted the importance of outreach to younger LGBTQIA+ individuals is seen as crucial for increasing diverse participation. One transgender man  from the US emphasized the need for reaching out to school students to show them potential career paths: \emph{“I feel like there should be some reaching out towards the younger community, like in high school or middle school, especially like LGBT groups, to show them this is a potential path for people who are queer. [...] Drop in and do like Makerspace or learn about computing or networking like they're just or even coding.”}

Supporting this view, a transgender woman from Canada also expressed her belief in engaging with schools: \emph{“I think we need to show young people, especially those from LGBTQIA+ communities, that we’re here and tech is a place for us too. We had a software group project here where we could invite schools directly, so we were able to show them what’s possible and inspire them to get involved, [...] just like anyone else.”} (P12)

\rqlearning{}{Inclusive course content, communities, outreach, representation, and \emph{consistent} support foster a welcoming environment.}

\subsubsection{Mixed Reactions to Exclusive Scholarships}\label{rq3-scholarships}
Participants expressed mixed feelings about scholarships and exclusive opportunities aimed at gender minorities. While some appreciated the targeted support, others felt uneasy about the exclusivity, fearing it might lead to tokenism or feelings of being singled out.

One participant appreciated scholarships targeting individuals like him: \emph{“I think scholarships towards people like me are more welcoming.”} (P10)

In contrast, one genderfluid student felt uneasy about scholarships exclusive to female students: \emph{“It felt more like people were being paraded more than actually being included.”} (P06)

While these opportunities can provide much-needed support to disadvantaged individuals, they must be handled delicately to avoid contributing to feelings of exclusion or tokenism.

\rqlearning{}{Scholarships for gender minority students are helpful but should be designed to avoid feelings of tokenism or marginalisation.}

\subsubsection{Structural Changes: Education and Training}
Participants (P01, P04, P06, P08, P11, P13) highlighted the need for structural changes, such as providing gender-neutral restrooms and raising awareness through training programs for staff and students.

One genderfluid student from Italy spoke about the difficulties faced by their friends regarding binary-gendered restrooms: \emph{“It would be helpful for those who don’t feel comfortable using male or female restrooms. Some of my friends really struggle with this.”}
In addition to practical changes, institutions can implement training programs to educate faculty, staff, and students about gender minorities. This can foster a more understanding and less judgmental environment, as stated: \emph{“Education about gender minorities could be beneficial for ensuring less judgment from others.”} (P13)

One participant also emphasised the potential of technology to build communities and improve access to resources for gender minorities: \emph{“There are so many things that would be better if everyone had better access to information. It would impact both women and gender non-conforming people, helping them access resources for their needs, including professional journeys and emotional support.”} (P06)

\rqlearning{}{Gender-neutrality in facilities and programs can enhance a more respectful and understanding academic environment.}

\subsubsection{Overcoming Gender}
Many students expressed a desire for a world where their gender identity no longer held significance in their academic or professional experiences. The ideal, as voiced by most participants, is that gender identity and expression should not matter at all.
Achieving this vision involves embracing neutrality to gender: \emph{“Perhaps 'indifference' sounds weird, but that's what I would like to see. Just people being indifferent to their gender like they. I would like to see people not actually caring so much. And yeah, just normalize it, you know.”} (P05)

This shift is not solely the responsibility of educational institutions but should be a broader societal effort. As another participant emphasised:
\emph{“I would say to not judge based on gender or based on how a person appears. Uh, that would be much more… better. Because I feel like uh, a lot of people judge based on their appearances, so if they appear as as a woman so they get treated just because of that or if they appear as a man, they get treated like a man. But. I would prefer if, if you would just treat them as a person and get interested in them as a person.”} (P04)
	
\rqsummary{RQ3}{The very basics are required to get more inclusive spaces: more representation, community support, gender-neutral facilities, and a shift in mindset, while avoiding tokenism.}

\section{Discussion}
Our findings reveal how software engineering education intersects with critical cultural, political, and societal issues.

\subsection{Thematic Overview and Interconnections}
\begin{figure*}
	\centering
	\includegraphics[width=1\linewidth]{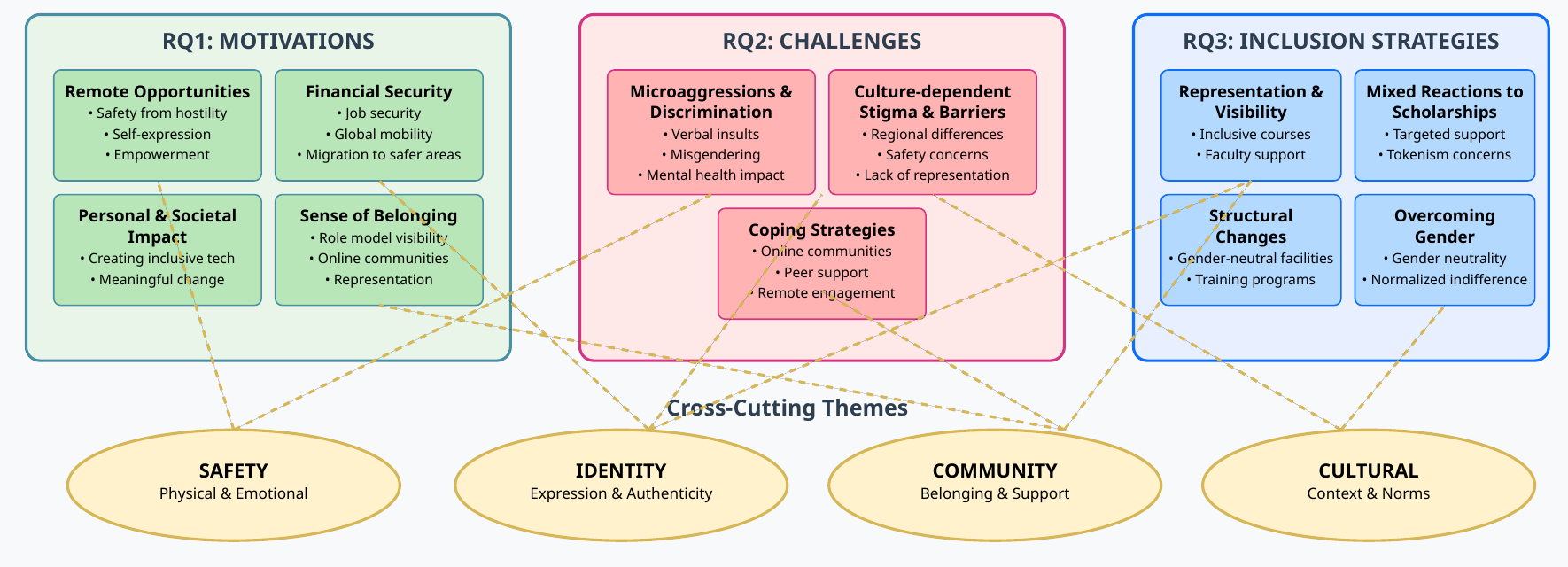}
	\caption{Thematic Map of Experiences of Transgender and Non-binary Students in Software Engineering. \small This figure presents key findings on our three research questions (RQ1--RQ3). The dotted lines connecting \emph{Safety}, \emph{Identity}, \emph{Community}, and \emph{Cultural} show that these four themes appear across all RQs, affecting motivations, challenges, and solutions alike.}
	\label{fig:overview}
\end{figure*}

\Cref{fig:overview} presents a thematic map that illustrates the relationships among our findings across all three research questions. Four cross-cutting themes emerge: safety (\emph{RQ1, RQ2}), identity (\emph{RQ1, RQ2, RQ3}), community (\emph{RQ1, RQ2, RQ3}), and cultural context (\emph{RQ2, RQ3}).

By excluding gender-minority students, we lose diverse perspectives essential for equitable software and teamwork~\cite{gardenswartz2003,earle2024,kohl2024}. Our participants' desire to create inclusive technologies reflects the need for software to increasingly shape critical aspects of life, yet it is often designed without input from marginalised communities~\cite{storey2020,betz2025}. The challenges reinforce workforce homogeneity, directly shaping which values are embedded in software~\cite{albusaysDiversityCrisisSoftware2021}. Many queer students view programming as a tool for social change, but recognise the field's role in perpetuating inequities due to under-representation~\cite{ryoo2024,vorderwulbeke2025}.
The global nature of our findings highlights how software engineering can either reinforce or challenge societal inequities. This is especially urgent given political shifts restricting transgender rights in some countries~\cite{desouzasantos2026} and AI systems that reinforce existing biases~\cite{hyrynsalmi2025c}.

\subsection{Societal and Educational Implications}
Educators play a critical role in fostering an inclusive environment for gender minorities in software engineering education.

\subsubsection{Remote Study and Work} 
Unlike many cisgender individuals who can typically choose remote work arrangements without significant concern, gender minority students often weigh the implications of their identity when making study and location choices (RQ1, RQ2). Offering flexible options, such as online courses or remote internships, can help mitigate these challenges~\cite{nolan2024}.
The preference for remote work aligns with findings from industry ~\cite{desouzasantosWhatTransgenderSoftware2023}. However, framing remote work as the primary solution risks normalising exclusion. If physical environments remain hostile, we create a two-tier system in which gender minorities must choose between safety and full participation, thereby affecting social integration, professional networking, and visibility~\cite {desouzasantosWhatTransgenderSoftware2023}.

\subsubsection{Course Materials}
Other STEM disciplines, such as biology, face similar challenges in integrating gender-inclusive perspectives, particularly in contexts where embracing diverse identities may encounter resistance \cite{lewis2024,harris2024}. 
However, given the role model of social sciences and humanities, simple changes can already make a difference~\cite{vorderwulbeke2025}.
For example, using \emph{they/them} pronouns and diverse gendered examples can help create a welcoming environment, as highlighted in RQ3. 
Course evaluations can include questions on gender inclusivity to provide improvement and consideration~\cite{call2023}.

\subsubsection{Faculty Training}
Despite the workload of most of the faculty staff, regular training on gender diversity and unconscious bias can provide actionable strategies for fostering inclusivity~\cite{harris2024,weissman2024}. This can be implemented as a brief annual workshop, or more innovative outreach methods, such as social media campaigns or reading groups~\cite{hopper2022, bakka2021}.

\subsubsection{Communities and Mentorship}
Creating gender-diverse spaces and mentorship programs that connect senior students with newcomers provides essential support~\cite{sulimani-aidan2024}, not only at universities but also at conferences for early PhD students. This could also result in LGBTQIA+ activities being included by \emph{default} at conferences, rather than being a \emph{nice-to-have}. To avoid tokenism (\Cref{rq3-scholarships}), programs should be open to all gender minorities or explicitly state target identities.

\subsubsection{Culture and Empathy}	
Cultural norms within educational institutions contribute to the exclusion of gender minorities (P03, P04, P05). Empathy and compassion are essential aspects in creating inclusive spaces, as the younger generation already recognises~\cite{ryoo2024}.
Empathy-driven interventions, such as (all year round) awareness campaigns, might shift norms to be more accepting and inclusive, as one student explains \emph{“Orientation matters. You know people come from different backgrounds. There are different understandings about the issues, that is just the problem, but as long as they can maybe try and I will create awareness where they can just get to understand that these people are also human being and we are... not something else.”} (P03)
Cultural differences also influence attitudes towards gender diversity globally, highlighting the need for context-specific solutions \cite{neumann2024}. 
While countries with higher income equality see greater inclusivity in online platforms, regions with higher levels of social aggression may present additional challenges \cite{zolduoarrati2024, mutongoza2024}.

\subsection{Implications for Research}
Our study aligns with findings indicating that more research is needed on this topic. We emphasise three approaches in software engineering research. 

\subsubsection{Outreach to Younger Students} 
As several participants noted (e.g., P10, P11), engaging younger students is essential to inspire them to pursue software engineering. Early exposure to the field, integrated into pre-university programs, could foster interest and create a more inclusive environment for gender minorities from the start~\cite{sulimani-aidan2024, ryoo2024}. 

\subsubsection{Post-Graduation Experiences} 
Many participants, such as P02, P05, and P10, voiced concerns about navigating future workplace environments and job applications, which may mirror the challenges faced in academic settings \cite{weissman2024}. Thus, investigating how these students fare after entering the workforce might provide insights into how (inclusive) educational practices translate into professional support, and which factors make a workplace a \emph{safe space} for them.

\subsubsection{Intersectionality}
Gender minority students are not a monolithic group~\cite{jennings2020}, and their experiences vary across different identities. The research community is encouraged to collect and report more fine-grained data, using an intersectional lens, to better understand and address the specific challenges faced by diverse students~\cite{garvey2019queer}: \emph{“I wish just there was more understanding of... Students who are different. Because I'll fulfill this also as a disabled person too. [...] People are just aren't aware that we exist.”} (P10)

\section{Conclusions}
In this study, we investigated the experiences of 13 transgender and non-binary students in software engineering. While some challenges are universal, students' experiences vary by cultural context, highlighting the importance of intersectional research that considers how gender identity intersects with socio-cultural dimensions. Our participants wished for more flexible study options, inclusive content, faculty training, and supportive communities. 

However, our findings reveal a deeper need: software engineering must give marginalised voices across education, research, and industry. As we build technologies that shape society's critical functions, excluding gender minorities embeds inequality into our digital future. Only by shifting the paradigm about diversity from an \emph{nice-to-have} attitude to truly incorporating marginalised voices into the software engineering process can we create technologies that serve all of society.

\begin{acks}
We sincerely thank the students for their openness and willingness to share their experiences. We are grateful to Ekin Su Gümüş for their initial impulse, assistance with recruitment and data analysis!
\end{acks}


\bibliographystyle{ACM-Reference-Format}
\bibliography{bibliography}

@article{albusaysDiversityCrisisSoftware2021,
  title = {The {{Diversity Crisis}} in {{Software Development}}},
  author = {Albusays, Khaled and Bjorn, Pernille and Dabbish, Laura and Ford, Denae and {Murphy-Hill}, Emerson and Serebrenik, Alexander and Storey, Margaret-Anne},
  year = {2021},
  month = mar,
  journal = {IEEE Software},
  volume = {38},
  number = {2},
  pages = {19--25},
  issn = {1937-4194},
  urldate = {2024-06-28}
}

@article{americanpsychologicalassociation2018,
  title = {Defining Transgender Terms},
  author = {American Psychological Association},
  year = {2018},
  journal = {Monitor on Psychology},
  series = {Vol 49},
  volume = {8},
  urldate = {2025-01-14},
  langid = {english}
}

@article{ashley2024,
  title = {Beyond the Trans/Cis Binary: Introducing New Terms Will Enrich Gender Research},
  shorttitle = {Beyond the Trans/Cis Binary},
  author = {Ashley, Florence and {Brightly-Brown}, Shari and Rider, G. Nic},
  year = {2024},
  month = jun,
  journal = {Nature},
  volume = {630},
  number = {8016},
  pages = {293--295},
  publisher = {Nature Publishing Group},
  urldate = {2025-01-14},
  copyright = {2024 Springer Nature Limited},
  langid = {english}
}

@article{atherton2016physics,
  title = {{{LGBT}} Climate in Physics: {{Building}} an Inclusive Community},
  author = {Atherton, Timothy J. and Barthelemy, Ram{\'O}n S. and Deconinck, Wouter and Falk, Michael L. and Garmon, Savannah and Long, Elena and Plisch, Monica and Simmons, Elizabeth H. and Reeves, Kyle},
  year = {2016},
  journal = {American Physical Society},
  publisher = {College Park}
}

@inproceedings{bakka2021,
  title = {Queering {{Engineering Through}} a {{Student Driven LGBTQIA}}+ {{Reading Group}} ({{Experience}})},
  booktitle = {{{ASEE}} Annual Conference Exposition},
  author = {Bakka, Brandon and Chou, Vivian and Marchioni, Jeffrey and Prince, Cassandra and Sugerman, Gabriella and Upreti, Ria and Clayton, Patricia and Borrego, Maura},
  year = {2021},
  urldate = {2025-01-16}
}

@article{baskararajah2021,
  title = {Term {{Interrelations}} and {{Trends}} in {{Software Engineering}}},
  author = {Baskararajah, Janusan and Zhang, Lei and Miranskyy, Andriy},
  year = {2021},
  month = aug,
  journal = {Proceedings of the 29th ACM Joint Meeting on European Software Engineering Conference and Symposium on the Foundations of Software Engineering},
  eprint = {2108.09529},
  pages = {1471--1474},
  urldate = {2022-03-04},
  archiveprefix = {arXiv}
}

@inproceedings{begoDiversityInclusionEngineering2021,
  title = {Diversity and Inclusion in Engineering and Computing: {{A}} Scoping Review of Recent {{FIE}} Papers},
  shorttitle = {Diversity and Inclusion in Engineering and Computing},
  booktitle = {2021 {{IEEE Frontiers}} in {{Education Conference}} ({{FIE}})},
  author = {Bego, Campbell R. and Nwokeji, Joshua C.},
  year = {2021},
  month = oct,
  pages = {1--9},
  issn = {2377-634X},
  urldate = {2024-06-28}
}

@article{beischel2023,
  title = {The Gender/Sex 3{\textbackslash}times 3: {{Measuring}} and Categorizing Gender/Sex beyond Binaries.},
  shorttitle = {The Gender/Sex 3{\textbackslash}times 3},
  author = {Beischel, Will J. and Schudson, Zach C. and Hoskin, Rhea Ashley and {van Anders}, Sari M.},
  year = {2023},
  journal = {Psychology of Sexual Orientation and Gender Diversity},
  volume = {10},
  number = {3},
  pages = {355},
  publisher = {Educational Publishing Foundation},
  urldate = {2025-01-14}
}

@article{bilimoria2009,
  title = {" {{Don}}'t Ask, Don't Tell": {{The}} Academic Climate for Lesbian, Gay, Bisexual, and Transgender Faculty in Science and Engineering},
  shorttitle = {" {{Don}}'t Ask, Don't Tell"},
  author = {Bilimoria, Diana and Stewart, Abigail J.},
  year = {2009},
  journal = {nwsa Journal},
  volume = {21},
  number = {2},
  pages = {85--103},
  publisher = {Johns Hopkins University Press},
  urldate = {2024-10-04}
}

@article{blincoe2019b,
  title = {Perceptions of {{Gender Diversity}}'s {{Impact}} on {{Mood}} in {{Software Development Teams}}},
  author = {Blincoe, Kelly and Springer, Olga and Wrobel, Michal R.},
  year = {2019},
  month = sep,
  journal = {IEEE Software},
  volume = {36},
  number = {5},
  pages = {51--56},
  issn = {1937-4194}
}

@inproceedings{boudreau2018,
  title = {Exploring Inclusive Spaces for {{LGBTQ}} Engineering Students},
  booktitle = {2018 {{CoNECD-The Collaborative Network}} for {{Engineering}} and {{Computing Diversity Conference}}},
  author = {Boudreau, Kristin and DiBiasio, David and Quinn, Paula and Reidinger, Zoe},
  year = {2018}
}

@book{braun2012,
  title = {Thematic Analysis.},
  author = {Braun, Virginia and Clarke, Victoria},
  year = {2012},
  publisher = {American Psychological Association},
  urldate = {2024-10-10}
}

@article{brinkworth2016chilly,
  title = {From Chilly Climate to Warm Reception: Experiences and Good Practices for Supporting {{LGBTQ}} Students in {{STEM}}},
  author = {Brinkworth, Carolyn S},
  year = {2016}
}

@article{call2023,
  title = {Transgender {{Inclusive}} and {{Affirming Design}} in {{Computing}}},
  author = {Call, Madison W and Roscoe, Rod D},
  year = {2023},
  journal = {Proceedings of the Human Factors and Ergonomics Society Annual Meeting},
  langid = {english}
}

@inproceedings{canedo2021c,
  title = {Breaking {{One Barrier}} at a {{Time}}: {{How Women Developers Cope}} in a {{Men-Dominated Industry}}},
  shorttitle = {Breaking {{One Barrier}} at a {{Time}}},
  booktitle = {Brazilian {{Symposium}} on {{Software Engineering}}},
  author = {Canedo, Edna Dias and Mendes, Fabiana and Cerqueira, Anderson and Okimoto, Marcio and Pinto, Gustavo and Bonifacio, Rodrigo},
  year = {2021},
  month = sep,
  pages = {378--387},
  publisher = {ACM},
  address = {Joinville Brazil},
  isbn = {978-1-4503-9061-3}
}

@inproceedings{canedo2023,
  title = {Do You See What Happens around You? {{Men}}'s {{Perceptions}} of {{Gender Inequality}} in {{Software Engineering}}},
  shorttitle = {Do You See What Happens around You?},
  booktitle = {Proceedings of the {{XXXVII Brazilian Symposium}} on {{Software Engineering}}},
  author = {Canedo, Edna Dias and Soares, Larissa and Silva, Geovana Ramos Sousa and Santos, Ver{\^o}nica Souza Dos and Mendes, Fabiana Freitas},
  year = {2023},
  month = sep,
  series = {{{SBES}} '23},
  pages = {464--474},
  publisher = {Association for Computing Machinery},
  address = {New York, NY, USA},
  urldate = {2023-10-23},
  isbn = {9798400707872}
}

@article{carver2021,
  title = {Behavioral {{Science}} and {{Diversity}} in {{Software Engineering}}},
  author = {Carver, J. C. and Muccini, H. and Penzenstadler, B. and Prikladnicki, R. and Serebrenik, A. and Zimmermann, T.},
  year = {2021},
  month = mar,
  journal = {IEEE Software},
  volume = {38},
  number = {2},
  pages = {107--112},
  issn = {1937-4194}
}

@article{casey2019,
  title = {Discrimination in the {{United States}}: {{Experiences}} of Lesbian, Gay, Bisexual, Transgender, and Queer {{Americans}}},
  shorttitle = {Discrimination in the {{United States}}},
  author = {Casey, Logan S. and Reisner, Sari L. and Findling, Mary G. and Blendon, Robert J. and Benson, John M. and Sayde, Justin M. and Miller, Carolyn},
  year = {2019},
  month = dec,
  journal = {Health Services Research},
  volume = {54},
  number = {S2},
  pages = {1454--1466},
  issn = {0017-9124, 1475-6773},
  urldate = {2025-01-14},
  langid = {english}
}

@article{casper2022revealing,
  title = {Revealing the Queer-Spectrum in {{STEM}} through Robust Demographic Data Collection in Undergraduate Engineering and Computer Science Courses at Four Institutions},
  author = {Casper, AM Aramati and Atadero, Rebecca A and Fuselier, Linda C},
  year = {2022},
  journal = {Plos one},
  volume = {17},
  number = {3},
  pages = {e0264267},
  publisher = {Public Library of Science San Francisco, CA USA}
}

@inproceedings{catolino2019,
  title = {Gender {{Diversity}} and {{Women}} in {{Software Teams}}: {{How Do They Affect Community Smells}}?},
  shorttitle = {Gender {{Diversity}} and {{Women}} in {{Software Teams}}},
  booktitle = {2019 {{IEEE}}/{{ACM}} 41st {{International Conference}} on {{Software Engineering}}: {{Software Engineering}} in {{Society}} ({{ICSE-SEIS}})},
  author = {Catolino, Gemma and Palomba, Fabio and Tamburri, Damian A. and Serebrenik, Alexander and Ferrucci, Filomena},
  year = {2019},
  month = may,
  pages = {11--20},
  publisher = {IEEE},
  address = {Montreal, QC, Canada},
  urldate = {2020-11-06},
  isbn = {978-1-72811-762-1},
  langid = {english}
}

@article{cheryan2024,
  title = {Global Patterns of Gender Disparities in {{STEM}} and Explanations for Their Persistence},
  author = {Cheryan, Sapna and Lombard, Ella J. and Hailu, Fasika and Pham, Linh N. H. and Weltzien, Katherine},
  year = {2024},
  month = nov,
  journal = {Nat Rev Psychol},
  pages = {1--14},
  publisher = {Nature Publishing Group},
  issn = {2731-0574},
  urldate = {2024-12-09},
  copyright = {2024 Springer Nature America, Inc.},
  langid = {english}
}

@misc{damian2024,
  title = {Equity, {{Diversity}}, and {{Inclusion}} in {{Software Engineering}}: {{Best Practices}} and {{Insights}}},
  shorttitle = {Equity, {{Diversity}}, and {{Inclusion}} in {{Software Engineering}}},
  author = {Damian, Daniela and Blincoe, Kelly and Ford, Denae and Serebrenik, Alexander and Masood, Zainab},
  year = {2024},
  publisher = {Springer Nature},
  urldate = {2025-01-16}
}

@inproceedings{desouzasantos2023,
  title = {Benefits and {{Limitations}} of {{Remote Work}} to {{LGBTQIA}}+ {{Software Professionals}}},
  booktitle = {2023 {{IEEE}}/{{ACM}} 45th {{International Conference}} on {{Software Engineering}}: {{Software Engineering}} in {{Society}} ({{ICSE-SEIS}})},
  author = {{de Souza Santos}, Ronnie and {de Magalh{\~a}es}, Cleyton V. C. and Ralph, Paul},
  year = {2023},
  month = may,
  pages = {48--57},
  issn = {2832-7616},
  urldate = {2024-10-07}
}

@inproceedings{desouzasantos2024,
  title = {Hidden {{Populations}} in {{Software Engineering}}: {{Challenges}}, {{Lessons Learned}}, and {{Opportunities}}},
  shorttitle = {Hidden {{Populations}} in {{Software Engineering}}},
  booktitle = {Proceedings of the 1st {{IEEE}}/{{ACM International Workshop}} on {{Methodological Issues}} with {{Empirical Studies}} in {{Software Engineering}}},
  author = {De Souza Santos, Ronnie and Gama, Kiev},
  year = {2024},
  month = apr,
  pages = {58--63},
  publisher = {ACM},
  address = {Lisbon Portugal},
  urldate = {2025-01-14},
  isbn = {9798400705670},
  langid = {english}
}

@inproceedings{desouzasantosDiversitySoftwareEngineering2023,
  title = {Diversity in Software Engineering: {{A}} Survey about Scientists from Underrepresented Groups},
  shorttitle = {Diversity in Software Engineering},
  booktitle = {2023 {{IEEE}}/{{ACM}} 16th {{International Conference}} on {{Cooperative}} and {{Human Aspects}} of {{Software Engineering}} ({{CHASE}})},
  author = {{de Souza Santos}, Ronnie and {Stuart-Verner}, Brody and {de Magalhaes}, Cleyton VC},
  year = {2023},
  pages = {161--166},
  publisher = {IEEE},
  urldate = {2024-06-10}
}

@article{desouzasantosWhatTransgenderSoftware2023,
  title = {What Do Transgender Software Professionals Say about a Career in the Software Industry?},
  author = {{de Souza Santos}, Ronnie and {Stuart-Verner}, Brody and Magalh{\~a}es, Cleyton},
  year = {2023},
  journal = {IEEE Software},
  publisher = {IEEE},
  urldate = {2024-06-11}
}

@inproceedings{earle2024,
  title = {Will {{I}} Fit? {{The}} Impact of Social and Identity Determinants on Teamwork in Engineering Education},
  shorttitle = {Will {{I}} Fit?},
  booktitle = {Frontiers in {{Education}}},
  author = {Earle, Shayna and McDonald, Madison and Bengizi, Esra and Jones, Kim S.},
  year = {2024},
  volume = {9},
  pages = {1412882},
  publisher = {Frontiers Media SA},
  urldate = {2025-01-16}
}

@article{earle2024a,
  title = {Will {{I}} Fit? {{The}} Impact of Social and Identity Determinants on Teamwork in Engineering Education},
  shorttitle = {Will {{I}} Fit?},
  author = {Earle, Shayna and McDonald, Madison and Bengizi, Esra and Jones, Kim},
  year = {2024},
  month = oct,
  journal = {Frontiers in Education},
  volume = {9}
}

@article{ellard-gray2015,
  title = {Finding the {{Hidden Participant}}: {{Solutions}} for {{Recruiting Hidden}}, {{Hard-to-Reach}}, and {{Vulnerable Populations}}},
  shorttitle = {Finding the {{Hidden Participant}}},
  author = {{Ellard-Gray}, Amy and Jeffrey, Nicole K. and Choubak, Melisa and Crann, Sara E.},
  year = {2015},
  month = dec,
  journal = {International Journal of Qualitative Methods},
  volume = {14},
  number = {5},
  pages = {1609406915621420},
  issn = {1609-4069, 1609-4069},
  urldate = {2024-10-10},
  langid = {english}
}

@article{ellis2009diversity,
  title = {Diversity and Inclusivity at University: {{A}} Survey of the Experiences of Lesbian, Gay, Bisexual and Trans ({{LGBT}}) Students in the {{UK}}},
  author = {Ellis, Sonja J},
  year = {2009},
  journal = {Higher Education},
  volume = {57},
  number = {6},
  pages = {723--739},
  publisher = {Springer}
}

@article{fitzgerald-russell2022,
  title = {Microaggression Experiences of Queer Science Students in Their Departments},
  author = {{Fitzgerald-Russell}, Madison L. and Kowalske, Megan Grunert},
  year = {2022},
  journal = {Journal of Research in Science, Mathematics and Technology Education},
  volume = {5},
  number = {2},
  pages = {131--153},
  publisher = {AX Publications},
  urldate = {2025-01-16}
}

@inproceedings{ford2019c,
  title = {How {{Remote Work Can Foster}} a {{More Inclusive Environment}} for {{Transgender Developers}}},
  booktitle = {2019 {{IEEE}}/{{ACM}} 2nd {{International Workshop}} on {{Gender Equality}} in {{Software Engineering}} ({{GE}})},
  author = {Ford, Denae and Milewicz, Reed and Serebrenik, Alexander},
  year = {2019},
  month = may,
  pages = {9--12},
  urldate = {2025-01-18}
}

@book{gardenswartz2003,
  title = {Diverse Teams at Work: {{Capitalizing}} on the Power of Diversity},
  shorttitle = {Diverse Teams at Work},
  author = {Gardenswartz, Lee and Rowe, Anita},
  year = {2003},
  publisher = {Society for Human Resource}
}

@article{garvey2018impact,
  title = {The Impact of Campus Climate on Queer-Spectrum Student Academic Success},
  author = {Garvey, Jason C and Squire, Dian D and Stachler, Brett and Rankin, Susan},
  year = {2018},
  journal = {Journal of LGBT Youth},
  volume = {15},
  number = {2},
  pages = {89--105},
  publisher = {Taylor \& Francis}
}

@article{garvey2019queer,
  title = {Queer and Trans* Students of Color: {{Navigating}} Identity Disclosure and College Contexts},
  author = {Garvey, Jason C and Mobley Jr, Steve D and Summerville, Kiara S and Moore, Gretchen T},
  year = {2019},
  journal = {The Journal of Higher Education},
  volume = {90},
  number = {1},
  pages = {150--178},
  publisher = {Taylor \& Francis}
}

@article{goldbergWantBeSafe2021,
  title = {``{{I Want}} to {{Be Safe}}{\dots}{{And I Also Want}} a {{Job}}'': {{Career Considerations}} and {{Decision-Making Among Transgender Graduate Students}}},
  shorttitle = {``{{I Want}} to {{Be Safe}}{\dots}{{And I Also Want}} a {{Job}}''},
  author = {Goldberg, Abbie E. and Matsuno, Emmie and Beemyn, Genny},
  year = {2021},
  month = nov,
  journal = {The Counseling Psychologist},
  volume = {49},
  number = {8},
  pages = {1147--1187},
  issn = {0011-0000, 1552-3861},
  urldate = {2024-06-11},
  langid = {english}
}

@article{gonzalez2024,
  title = {'{{Whenever I}} See Those Little Rainbow Stickers, {{I}} Know That There Is a Place You Can Go': {{Visibility}} and Sense of Belonging for Queer and/or Trans Community College Students},
  shorttitle = {'{{Whenever I}} See Those Little Rainbow Stickers, {{I}} Know That There Is a Place You Can Go'},
  author = {Gonz{\'a}lez, {\'A}ngel de Jesus},
  year = {2024},
  month = may,
  journal = {Journal of Diversity in Higher Education},
  publisher = {Educational Publishing Foundation},
  issn = {1938-8926},
  urldate = {2024-10-04}
}

@article{groeneveld2022a,
  title = {Identifying {{Non-Technical Skill Gaps}} in {{Software Engineering Education}}: {{What Experts Expect But Students Don}}'t {{Learn}}},
  shorttitle = {Identifying {{Non-Technical Skill Gaps}} in {{Software Engineering Education}}},
  author = {Groeneveld, Wouter and Vennekens, Joost and Aerts, Kris},
  year = {2022},
  month = mar,
  journal = {ACM Transactions on Computing Education},
  volume = {22},
  pages = {1--21}
}

@article{harris2024,
  title = {Incorporating {{Sex-Diverse}} and {{Gender-Inclusive Perspectives}} in {{Higher Education Biology Courses}}},
  author = {Harris, Breanna N and Lewis, A Kelsey and Sharpe, Sam L and Orr, Teri J and Martine, Christopher T and Josefson, Chloe C},
  year = {2024},
  month = dec,
  journal = {Integrative and Comparative Biology},
  volume = {64},
  number = {6},
  pages = {1694--1716},
  issn = {1540-7063},
  urldate = {2025-01-16}
}

@inproceedings{hopper2022,
  title = {The Implementation and Assessment of a Social Media Initiative to Increase Visibility of {{LGBTQIA}}+ Individuals in {{STEM}}},
  booktitle = {2022 {{ASEE Annual Conference}} \& {{Exposition}}},
  author = {Hopper, Theo S. and {Tossas-Betancourt}, Christopher and Walczyk, Peter and Hirshfield, Laura},
  year = {2022},
  urldate = {2025-01-16}
}

@inproceedings{jennings2020,
  title = {A {{Review}} of the {{State}} of {{LGBTQIA}}+ {{Student Research}} in {{STEM}} and {{Engineering Education}}},
  booktitle = {2020 {{ASEE Virtual Annual Conference Content Access Proceedings}}},
  author = {Jennings, Madeleine and Roscoe, Rod and Kellam, Nadia and Jayasuriya, Suren},
  year = {2020},
  month = jun,
  pages = {34045},
  publisher = {ASEE Conferences},
  address = {Virtual On line},
  urldate = {2024-10-04},
  langid = {english}
}

@article{jourian2017,
  title = {Bringing Our Communities to the Research Table: The Liberatory Potential of Collaborative Methodological Practices alongside {{LGBTQ}} Participants},
  shorttitle = {Bringing Our Communities to the Research Table},
  author = {Jourian, T.J. and Nicolazzo, Z},
  year = {2017},
  month = aug,
  journal = {Educational Action Research},
  volume = {25},
  number = {4},
  pages = {594--609},
  issn = {0965-0792, 1747-5074},
  urldate = {2025-01-16},
  langid = {english}
}

@article{kidd2021,
  title = {Prevalence of {{Gender-Diverse Youth}} in an {{Urban School District}}},
  author = {Kidd, Kacie M. and Sequeira, Gina M. and Douglas, Claudia and Paglisotti, Taylor and {Inwards-Breland}, David J. and Miller, Elizabeth and Coulter, Robert W. S.},
  year = {2021},
  month = jun,
  journal = {Pediatrics},
  volume = {147},
  number = {6},
  pages = {e2020049823},
  issn = {0031-4005, 1098-4275},
  urldate = {2025-01-14},
  langid = {english}
}

@incollection{kohl2024,
  title = {Gender {{Diversity}} on {{Software Development Teams}}: {{A Qualitative Study}}},
  shorttitle = {Gender {{Diversity}} on {{Software Development Teams}}},
  booktitle = {Equity, {{Diversity}}, and {{Inclusion}} in {{Software Engineering}}: {{Best Practices}} and {{Insights}}},
  author = {Kohl, Karina and Prikladnicki, Rafael},
  editor = {Damian, Daniela and Blincoe, Kelly and Ford, Denae and Serebrenik, Alexander and Masood, Zainab},
  year = {2024},
  pages = {169--184},
  publisher = {Apress},
  address = {Berkeley, CA},
  urldate = {2025-01-18},
  isbn = {978-1-4842-9651-6},
  langid = {english}
}

@article{kosciw2015,
  title = {Reflecting {{Resiliency}}: {{Openness About Sexual Orientation}} and/or {{Gender Identity}} and {{Its Relationship}} to {{Well}}-{{Being}} and {{Educational Outcomes}} for {{LGBT Students}}},
  shorttitle = {Reflecting {{Resiliency}}},
  author = {Kosciw, Joseph G. and Palmer, Neal A. and Kull, Ryan M.},
  year = {2015},
  month = mar,
  journal = {American J of Comm Psychol},
  volume = {55},
  number = {1-2},
  pages = {167--178},
  issn = {0091-0562, 1573-2770},
  urldate = {2024-10-09},
  copyright = {http://onlinelibrary.wiley.com/termsAndConditions\#vor},
  langid = {english}
}

@article{lewis2024,
  title = {Let's Talk about Sex: Instructor Views and Hesitancies Related to Sex and Gender in the Biology Classroom},
  shorttitle = {Let's Talk about Sex},
  author = {Lewis, A. Kelsey and Josefson, Chloe C. and Orr, Teri J. and Harris, Breanna N.},
  year = {2024},
  journal = {Integrative and Comparative Biology},
  volume = {64},
  number = {6},
  pages = {1679--1693},
  publisher = {Oxford University Press},
  urldate = {2025-01-16}
}

@article{lombardi2002,
  title = {Gender {{Violence}}: {{Transgender Experiences}} with {{Violence}} and {{Discrimination}}},
  shorttitle = {Gender {{Violence}}},
  author = {Lombardi, Emilia L. and Wilchins, Riki Anne and Priesing, Dana and Malouf, Diana},
  year = {2002},
  month = mar,
  journal = {Journal of Homosexuality},
  volume = {42},
  number = {1},
  pages = {89--101},
  issn = {0091-8369, 1540-3602},
  urldate = {2025-01-14},
  langid = {english}
}

@article{maloyFactorsInfluencingRetention2022,
  title = {Factors {{Influencing Retention}} of {{Transgender}} and {{Gender Nonconforming Students}} in {{Undergraduate STEM Majors}}},
  author = {Maloy, Jeffrey and Kwapisz, Monika B. and Hughes, Bryce E.},
  year = {2022},
  journal = {CBE Life Sciences Education},
  volume = {21},
  number = {1},
  pages = {ar13},
  issn = {1931-7913},
  urldate = {2024-06-27},
  pmcid = {PMC9250371},
  pmid = {35044846}
}

@article{mara2021strategies,
  title = {Strategies for Coping with {{LGBT}} Discrimination at Work: {{A}} Systematic Literature Review},
  author = {Mara, Liviu-Catalin and Ginieis, Mat{\'{\i}}as and {Brunet-Icart}, Ignasi},
  year = {2021},
  journal = {Sexuality Research and Social Policy},
  volume = {18},
  pages = {339--354},
  publisher = {Springer}
}

@article{mchugh2012,
  title = {Interrater Reliability: The Kappa Statistic},
  shorttitle = {Interrater Reliability},
  author = {McHugh, Mary L.},
  year = {2012},
  month = oct,
  journal = {Biochem Med (Zagreb)},
  volume = {22},
  number = {3},
  pages = {276--282},
  issn = {1330-0962},
  urldate = {2024-10-10},
  pmcid = {PMC3900052},
  pmid = {23092060}
}

@inproceedings{menierBroadeningGenderComputing2021,
  title = {Broadening Gender in Computing for Transgender and Nonbinary Learners},
  booktitle = {2021 {{Conference}} on {{Research}} in {{Equitable}} and {{Sustained Participation}} in {{Engineering}}, {{Computing}}, and {{Technology}} ({{RESPECT}})},
  author = {Menier, Amanda and Zarch, Rebecca and Sexton, Stacy},
  year = {2021},
  pages = {1--5},
  publisher = {IEEE},
  urldate = {2024-06-11}
}

@article{miller2021,
  title = {'{{It}}'s Dude Culture': {{Students}} with Minoritized Identities of Sexuality and/or Gender Navigating {{STEM}} Majors},
  shorttitle = {'{{It}}'s Dude Culture'},
  author = {Miller, Ryan A. and Vaccaro, Annemarie and Kimball, Ezekiel W. and Forester, Rachael},
  year = {2021},
  month = sep,
  journal = {Journal of Diversity in Higher Education},
  volume = {14},
  number = {3},
  pages = {340--352},
  publisher = {Educational Publishing Foundation},
  issn = {1938-8926},
  urldate = {2024-10-04}
}

@article{mutongoza2024,
  title = {Exploring the {{Psychosocial Outcomes}} of {{Microaggressions}} against {{Queer Communities}} at a {{Rural University}} in {{South Africa}}},
  author = {Mutongoza, Bonginkosi Hardy},
  year = {2024},
  journal = {The International Journal of Diverse Identities},
  volume = {24},
  number = {1},
  pages = {1--18},
  issn = {2327-7866, 2327-8560},
  urldate = {2025-01-16},
  langid = {english}
}

@incollection{neuendorfContentAnalysisThematic2018,
  title = {Content Analysis and Thematic Analysis},
  booktitle = {Advanced Research Methods for Applied Psychology},
  author = {Neuendorf, Kimberly A.},
  year = {2018},
  pages = {211--223},
  publisher = {Routledge},
  urldate = {2024-07-11}
}

@misc{neumann2024,
  title = {What {{You Use}} Is {{What You Get}}: {{Unforced Errors}} in {{Studying Cultural Aspects}} in {{Agile Software Development}}},
  shorttitle = {What {{You Use}} Is {{What You Get}}},
  author = {Neumann, Michael and Schmid, Klaus and Baumann, Lars},
  year = {2024},
  month = apr,
  number = {arXiv:2404.17009},
  eprint = {2404.17009},
  primaryclass = {cs},
  publisher = {arXiv},
  urldate = {2025-01-15},
  archiveprefix = {arXiv},
  langid = {english}
}

@misc{nolan2024,
  title = {Behind {{Closed Doors}}: {{The Untold Challenges}} of {{Transgender}} and {{Nonbinary Graduate Students}} in {{Chemistry}}},
  shorttitle = {Behind {{Closed Doors}}},
  author = {Nolan, Michelle M. and Blythe, Isaac M. and {Vincent-Ruz}, Paulette},
  year = {2024},
  month = may,
  urldate = {2025-01-16},
  copyright = {https://creativecommons.org/licenses/by-nc-nd/4.0/},
  langid = {english}
}

@article{oconnor2020,
  title = {Intercoder {{Reliability}} in {{Qualitative Research}}: {{Debates}} and {{Practical Guidelines}}},
  shorttitle = {Intercoder {{Reliability}} in {{Qualitative Research}}},
  author = {O'Connor, Cliodhna and Joffe, Helene},
  year = {2020},
  month = jan,
  journal = {International Journal of Qualitative Methods},
  volume = {19},
  pages = {1609406919899220},
  issn = {1609-4069, 1609-4069},
  urldate = {2025-01-14},
  langid = {english}
}

@inproceedings{pournaghshbandPromotingDiversityInclusiveComputer2020,
  title = {Promoting {{Diversity-Inclusive Computer Science Pedagogies}}: {{A Multidimensional Perspective}}},
  shorttitle = {Promoting {{Diversity-Inclusive Computer Science Pedagogies}}},
  booktitle = {Proceedings of the 2020 {{ACM Conference}} on {{Innovation}} and {{Technology}} in {{Computer Science Education}}},
  author = {Pournaghshband, Vahab and Medel, Paola},
  year = {2020},
  month = jun,
  pages = {219--224},
  publisher = {ACM},
  address = {Trondheim Norway},
  urldate = {2024-06-28},
  isbn = {978-1-4503-6874-2},
  langid = {english}
}

@article{pradell2024,
  title = {The Identity-Related Experiences of {{LGBTQ}} + Students in Engineering Spaces},
  author = {Pradell, Lee R. and Parmenter, Joshua G. and Galliher, Renee V. and Berke, Ryan B. and Rowley, Lindsey},
  year = {2024},
  month = sep,
  journal = {International Journal of Qualitative Studies in Education},
  volume = {37},
  number = {8},
  pages = {2267--2287},
  issn = {0951-8398, 1366-5898},
  urldate = {2025-01-16},
  langid = {english}
}

@article{prado2020,
  title = {How Trans-Inclusive Are Hackathons?},
  author = {Prado, Rafa and Mendes, Wendy and Gama, Kiev S. and Pinto, Gustavo},
  year = {2020},
  journal = {IEEE Software},
  volume = {38},
  number = {2},
  pages = {26--31},
  publisher = {IEEE},
  urldate = {2025-01-14}
}

@misc{ralph2021,
  title = {Empirical {{Standards}} for {{Software Engineering Research}}},
  author = {Ralph, Paul and bin Ali, Nauman and Baltes, Sebastian and Bianculli, Domenico and Diaz, Jessica and Dittrich, Yvonne and Ernst, Neil and Felderer, Michael and Feldt, Robert and Filieri, Antonio and de Fran{\c c}a, Breno Bernard Nicolau and Furia, Carlo Alberto and Gay, Greg and Gold, Nicolas and Graziotin, Daniel and He, Pinjia and Hoda, Rashina and Juristo, Natalia and Kitchenham, Barbara and Lenarduzzi, Valentina and Mart{\'i}nez, Jorge and Melegati, Jorge and Mendez, Daniel and Menzies, Tim and Molleri, Jefferson and Pfahl, Dietmar and Robbes, Romain and Russo, Daniel and Saarim{\"a}ki, Nyyti and Sarro, Federica and Taibi, Davide and Siegmund, Janet and Spinellis, Diomidis and Staron, Miroslaw and Stol, Klaas and Storey, Margaret-Anne and Taibi, Davide and Tamburri, Damian and Torchiano, Marco and Treude, Christoph and Turhan, Burak and Wang, Xiaofeng and Vegas, Sira},
  year = {2021},
  month = mar,
  number = {arXiv:2010.03525},
  eprint = {2010.03525},
  primaryclass = {cs},
  publisher = {arXiv},
  urldate = {2025-01-15},
  archiveprefix = {arXiv}
}

@article{reggianiLGBTAcademicsPhD2023,
  title = {{{LGBT}} + Academics' and {{PhD}} Students' Experiences of Visibility in {{STEM}}: {{More}} than Raising the Rainbow Flag},
  shorttitle = {{{LGBT}} + Academics' and {{PhD}} Students' Experiences of Visibility in {{STEM}}},
  author = {Reggiani, Marco and Gagnon, Jessica Dawn and Lunn, Rebecca Jane},
  year = {2023},
  month = jan,
  journal = {Higher Education},
  issn = {0018-1560, 1573-174X},
  urldate = {2024-06-28},
  langid = {english}
}

@article{rodriguez-perez2021d,
  title = {Perceived Diversity in Software Engineering: A Systematic Literature Review},
  shorttitle = {Perceived Diversity in Software Engineering},
  author = {{Rodr{\'i}guez-P{\'e}rez}, Gema and Nadri, Reza and Nagappan, Meiyappan},
  year = {2021},
  month = sep,
  journal = {Empir Software Eng},
  volume = {26},
  number = {5},
  pages = {102},
  issn = {1382-3256, 1573-7616},
  urldate = {2025-01-14},
  langid = {english}
}

@article{rosemarypradell2024,
  title = {{{LGBTQ}} + Engineering Students' Recommendations for Sustaining and Supporting Diversity in {{STEM}}},
  author = {Rosemary Pradell, Lee and Parmenter, Joshua G. and Galliher, Renee V. and Berke, Ryan},
  year = {2024},
  month = feb,
  journal = {Journal of LGBT Youth},
  pages = {1--31},
  issn = {1936-1653, 1936-1661},
  urldate = {2025-01-16},
  langid = {english}
}

@misc{russo2022b,
  title = {Recruiting {{Software Engineers}} on {{Prolific}}},
  author = {Russo, Daniel},
  year = {2022},
  month = mar,
  number = {arXiv:2203.14695},
  eprint = {2203.14695},
  primaryclass = {cs},
  publisher = {arXiv},
  urldate = {2022-07-05},
  archiveprefix = {arXiv},
  langid = {english}
}

@article{ryoo2024,
  title = {``{{Show}} Them the Playbook That These Companies Are Using'': {{Youth Voices}} about Why {{Computer Science Education Must Center Discussions}} of {{Power}}, {{Ethics}}, and {{Culturally Responsive Computing}}},
  shorttitle = {``{{Show}} Them the Playbook That These Companies Are Using''},
  author = {Ryoo, Jean J. and Blunt, Takeria},
  year = {2024},
  month = apr,
  journal = {ACM Trans. Comput. Educ.},
  pages = {3660645},
  issn = {1946-6226},
  urldate = {2024-05-21},
  langid = {english}
}

@misc{santosLGBTQIAVisibilityComputer2023,
  title = {{{LGBTQIA}}+ ({{In}}){{Visibility}} in {{Computer Science}} and {{Software Engineering Education}}},
  author = {Santos, Ronnie de Souza and {Stuart-Verner}, Brody and {de Magalhaes}, Cleyton},
  year = {2023},
  month = mar,
  number = {arXiv:2303.05953},
  eprint = {2303.05953},
  primaryclass = {cs},
  publisher = {arXiv},
  urldate = {2024-06-10},
  archiveprefix = {arXiv}
}

@article{sarna2021,
  title = {The {{Importance}} of {{Mentors}} and {{Mentoring Programs}} for {{LGBT}}+{{Undergraduate Students}}},
  author = {Sarna, Vincent and Dentato, Michael P. and DiClemente, Cara M. and Richards, Maryse H.},
  year = {2021},
  journal = {College Student Affairs Journal},
  volume = {39},
  number = {2},
  pages = {180--199},
  issn = {2381-2338},
  urldate = {2024-06-03},
  langid = {english}
}

@article{sona2023,
  title = {Looking through a {{Prism}}: {{A Systematic Review}} of {{LGBTQ}}+ {{STEM Literature}}},
  shorttitle = {Looking through a {{Prism}}},
  author = {Sona, Aj and Laboy Santana, Jabdiel and Saitta, Erin K.H.},
  year = {2023},
  month = jan,
  journal = {J. Chem. Educ.},
  volume = {100},
  number = {1},
  pages = {125--133},
  issn = {0021-9584, 1938-1328},
  urldate = {2025-01-16},
  copyright = {https://doi.org/10.15223/policy-029},
  langid = {english}
}

@article{staron2024a,
  title = {Laws, {{Ethics}}, and {{Fairness}} in {{Software Engineering}}},
  author = {Staron, Miroslaw and Abrah{\'a}o, Silvia and Serebrenik, Alexander and Penzenstadler, Birgit and Horkoff, Jennifer and Honnenahalli, Chetan},
  year = {2024},
  journal = {IEEE Software},
  volume = {42},
  number = {1},
  pages = {110--113},
  publisher = {IEEE},
  urldate = {2025-01-16}
}

@article{storey2020,
  title = {The {{Who}}, {{What}}, {{How}} of {{Software Engineering Research}}: {{A Socio-Technical Framework}}},
  shorttitle = {The {{Who}}, {{What}}, {{How}} of {{Software Engineering Research}}},
  author = {Storey, Margaret-Anne and Ernst, Neil A. and Williams, Courtney and Kalliamvakou, Eirini},
  year = {2020},
  month = may,
  journal = {arXiv:1905.12841 [cs]},
  eprint = {1905.12841},
  primaryclass = {cs},
  urldate = {2020-11-06},
  archiveprefix = {arXiv},
  langid = {english}
}

@article{stoutLesbianGayBisexual2016a,
  title = {Lesbian, {{Gay}}, {{Bisexual}}, {{Transgender}}, and {{Queer Students}}' {{Sense}} of {{Belonging}} in {{Computing}}: {{An Intersectional Approach}}},
  shorttitle = {Lesbian, {{Gay}}, {{Bisexual}}, {{Transgender}}, and {{Queer Students}}' {{Sense}} of {{Belonging}} in {{Computing}}},
  author = {Stout, Jane G. and Wright, Heather M.},
  year = {2016},
  month = may,
  journal = {Computing in Science \& Engineering},
  volume = {18},
  number = {3},
  pages = {24--30},
  issn = {1558-366X},
  urldate = {2024-06-27}
}

@article{sulimani-aidan2024,
  title = {Increasing Resilience among {{LGBTQ}} Youth: {{The}} Protective Role of Natural Mentors},
  shorttitle = {Increasing Resilience among {{LGBTQ}} Youth},
  author = {{Sulimani-Aidan}, Yafit and Shilo, Guy and Paul, June C.},
  year = {2024},
  month = may,
  journal = {Children and Youth Services Review},
  volume = {160},
  pages = {107570},
  issn = {0190-7409},
  urldate = {2024-06-03}
}

@article{syed2015,
  title = {Guidelines for {{Establishing Reliability When Coding Narrative Data}}},
  author = {Syed, Moin and Nelson, Sarah C.},
  year = {2015},
  month = dec,
  journal = {Emerging Adulthood},
  volume = {3},
  number = {6},
  pages = {375--387},
  issn = {2167-6968, 2167-6984},
  urldate = {2025-01-14},
  langid = {english}
}

@article{tracy2010,
  title = {Qualitative {{Quality}}: {{Eight}} ``{{Big-Tent}}'' {{Criteria}} for {{Excellent Qualitative Research}}},
  shorttitle = {Qualitative {{Quality}}},
  author = {Tracy, Sarah J.},
  year = {2010},
  month = dec,
  journal = {Qualitative Inquiry},
  volume = {16},
  number = {10},
  pages = {837--851},
  issn = {1077-8004, 1552-7565},
  urldate = {2025-01-14},
  copyright = {https://journals.sagepub.com/page/policies/text-and-data-mining-license},
  langid = {english}
}

@inproceedings{vanbreukelen2023a,
  title = {``{{STILL AROUND}}'': {{Experiences}} and {{Survival Strategies}} of {{Veteran Women Software Developers}}},
  shorttitle = {``{{STILL AROUND}}''},
  booktitle = {2023 {{IEEE}}/{{ACM}} 45th {{International Conference}} on {{Software Engineering}} ({{ICSE}})},
  author = {Van Breukelen, Sterre and Barcombt, Ann and Baltes, Sebastian and Serebrenik, Alexander},
  year = {2023},
  month = may,
  pages = {1148--1160},
  issn = {1558-1225},
  urldate = {2024-10-11}
}

@incollection{vinson2008,
  title = {A {{Practical Guide}} to {{Ethical Research Involving Humans}}},
  booktitle = {Guide to {{Advanced Empirical Software Engineering}}},
  author = {Vinson, Norman G. and Singer, Janice},
  editor = {Shull, Forrest and Singer, Janice and Sj{\o}berg, Dag I. K.},
  year = {2008},
  pages = {229--256},
  publisher = {Springer},
  address = {London},
  urldate = {2020-10-28},
  isbn = {978-1-84800-044-5},
  langid = {english}
}

@article{weissman2024,
  title = {Running a Queer-and Trans-Inclusive Faculty Hiring Process},
  author = {Weissman, J. L. and Chappell, Callie R. and {de Oliveira}, Bruno Francesco Rodrigues and Evans, Natalya and Fagre, Anna C. and Forsythe, Desiree and Frese, Steven A. and Gregor, Rachel and Ishaq, Suzanne L. and Johnston, Julie},
  year = {2024},
  publisher = {EcoEvoRxiv},
  urldate = {2025-01-16}
}

@article{whitley2022,
  title = {I've {{Been Misgendered So Many Times}}: {{Comparing}} the {{Experiences}} of {{Chronic Misgendering}} among {{Transgender Graduate Students}} in the {{Social}} and {{Natural Sciences}}},
  shorttitle = {I've {{Been Misgendered So Many Times}}},
  author = {Whitley, Cameron T. and Nordmarken, Sonny and Kolysh, Simone and Goldstein-Kral, Jess},
  year = {2022},
  month = aug,
  journal = {Sociological Inquiry},
  volume = {92},
  number = {3},
  pages = {1001--1028},
  issn = {0038-0245, 1475-682X},
  urldate = {2025-01-16},
  langid = {english}
}

@book{wohlin2012,
  title = {Experimentation in {{Software Engineering}}},
  author = {Wohlin, Claes and Runeson, Per and H{\"o}st, Martin and Ohlsson, Magnus C. and Regnell, Bj{\"o}rn and Wessl{\'e}n, Anders},
  year = {2012},
  publisher = {Springer-Verlag},
  address = {Berlin Heidelberg},
  urldate = {2019-10-29},
  isbn = {978-3-642-29043-5},
  langid = {english}
}

@article{yang2021,
  title = {{{RESISTANCE AND COMMUNITY-BUILDING IN LGBTQ}}+ {{ENGINEERING STUDENTS}}},
  author = {Yang, Jerry A. and Sherard, Max K. and Julien, Christine and Borrego, Maura},
  year = {2021},
  journal = {J Women Minor Scien Eng},
  volume = {27},
  number = {4},
  pages = {1--33},
  issn = {1072-8325},
  urldate = {2025-01-16},
  langid = {english}
}

@article{zolduoarrati2024,
  title = {Harmonising {{Contributions}}: {{Exploring Diversity}} in {{Software Engineering}} through {{CQA Mining}} on {{Stack Overflow}}},
  shorttitle = {Harmonising {{Contributions}}},
  author = {Zolduoarrati, Elijah and Licorish, Sherlock A. and Stanger, Nigel},
  year = {2024},
  month = jun,
  journal = {ACM Trans. Softw. Eng. Methodol.},
  issn = {1049-331X},
  urldate = {2024-07-03},
}

@article{landis1977,
  title={The measurement of observer agreement for categorical data},
  author={Landis, J Richard and Koch, Gary G},
  journal={biometrics},
  pages={159--174},
  year={1977},
  publisher={JSTOR}
}

@article{herman2025,
  title = {How Many Adults and Youth Identify as Transgender in the {{United States}}?},
  author = {Herman, Jody L},
  year = {2025},
  month = aug,
  langid = {english}
}

@inproceedings{vorderwulbeke2025,
  title={Belonging Beyond Code: Queer Software Engineering and Humanities Student Experiences},
  author={Vorderw{\"u}lbeke, Emily and Gra{\ss}l, Isabella},
  booktitle={2025 IEEE/ACM 47th International Conference on Software Engineering: Software Engineering in Society (ICSE-SEIS)},
  pages={13--24},
  year={2025},
  organization={IEEE}
}

@article{betz2025,
  title = {With {{Great Power Comes Great Responsibility}}: {{The Role}} of {{Software Engineers}}},
  shorttitle = {With {{Great Power Comes Great Responsibility}}},
  author = {Betz, Stefanie and Penzenstadler, Birgit},
  year = {2025},
  month = jun,
  journal = {ACM Transactions on Software Engineering and Methodology},
  volume = {34},
  number = {5},
  pages = {1--21},
  issn = {1049-331X, 1557-7392},
  urldate = {2025-09-29},
  langid = {english}
}

@incollection{desouzasantos2026,
  title = {From {{Diverse Origins}} to a {{DEI Crisis}}: {{The Pushback Against Equity}}, {{Diversity}}, and {{Inclusion}} in {{Software Engineering}}},
  shorttitle = {From {{Diverse Origins}} to a {{DEI Crisis}}},
  booktitle = {Software {{Engineering}} and {{Advanced Applications}}},
  author = {De Souza Santos, Ronnie and Magalhaes, Cleyton and Barcomb, Ann and Wessel, Mairieli},
  editor = {Taibi, Davide and Smite, Darja},
  year = {2026},
  volume = {16083},
  pages = {174--190},
  publisher = {Springer Nature Switzerland},
  address = {Cham},
  urldate = {2025-09-29},
  isbn = {978-3-032-04206-4 978-3-032-04207-1},
  langid = {english}
}

@article{hyrynsalmi2025c,
  title = {Making {{Software Development More Diverse}} and {{Inclusive}}: {{Key Themes}}, {{Challenges}}, and {{Future Directions}}},
  shorttitle = {Making {{Software Development More Diverse}} and {{Inclusive}}},
  author = {Hyrynsalmi, Sonja M. and Baltes, Sebastian and Brown, Chris and Prikladnicki, Rafael and {Rodriguez-Perez}, Gema and Serebrenik, Alexander and Simmonds, Jocelyn and Trinkenreich, Bianca and Wang, Yi and Liebel, Grischa},
  year = {2025},
  month = jun,
  journal = {ACM Transactions on Software Engineering and Methodology},
  volume = {34},
  number = {5},
  pages = {1--23},
  issn = {1049-331X, 1557-7392},
  urldate = {2025-09-29},
  langid = {english}
}
\end{document}